\documentclass[twocolumn]{aastex62}

\usepackage{booktabs} %Handling tables
\usepackage{multirow} %Handling tables
\usepackage{makecell} %Handling tables
\usepackage{comment}
\usepackage{xcolor} %Colour text

\newcommand{\Mj}{\rm M_{Jup}}

\newcommand{\hd}{HD\,163296}

\newcommand{\discminer}{\textsc{discminer}}
\newcommand{\twCOfull}{$^{12}$CO\,$J=2-1$}
\newcommand{\twCO}{$^{12}$CO}
\newcommand{\emcee}{\textsc{emcee}}
\defcitealias{izquierdo+2021}{Paper I}

\graphicspath{{./}{figures/}}

\received{June 15, 2021}
\revised{--}
\accepted{December 29, 2021}
\shorttitle{Mining the disc of \hd{}}
\shortauthors{Izquierdo et al.}
\begin{document}

\title{\textbf{A new planet candidate detected in a dust gap of the disc around HD\,163296 through localised kinematic signatures}\\An observational validation of the \discminer{}}

\correspondingauthor{Andr\'es F. Izquierdo}
\email{andres.izquierdo.c@gmail.com}

\author[0000-0001-8446-3026]{Andr\'es F. Izquierdo}
\affiliation{European Southern Observatory, Karl-Schwarzschild-Str. 2, 85748 Garching bei München, Germany}
\affiliation{Leiden Observatory, Leiden University, P.O. Box 9513, NL-2300 RA Leiden, The Netherlands}

\author[0000-0003-4689-2684]{Stefano Facchini}
\affiliation{European Southern Observatory, Karl-Schwarzschild-Str. 2, 85748 Garching bei München, Germany}

\author[0000-0003-4853-5736]{Giovanni P. Rosotti}
\affiliation{School of Physics and Astronomy, University of Leicester, Leicester LE1 7RH, UK}
\affiliation{Leiden Observatory, Leiden University, P.O. Box 9513, NL-2300 RA Leiden, The Netherlands}

\author[0000-0001-7591-1907]{Ewine F. van Dishoeck}
\affiliation{Leiden Observatory, Leiden University, P.O. Box 9513, NL-2300 RA Leiden, The Netherlands}
\affiliation{Max-Planck-Institut für extraterrestrische Physik, Gießenbachstr. 1 , 85748 Garching bei München, Germany}

\author[0000-0003-1859-3070]{Leonardo Testi}
\affiliation{European Southern Observatory, Karl-Schwarzschild-Str. 2, 85748 Garching bei München, Germany}
\affiliation{INAF – Osservatorio Astrofisico di Arcetri, Largo E. Fermi 5, 50125 Firenze, Italy}



\begin{abstract}
We report the robust detection of coherent, localised deviations from Keplerian rotation possibly associated with the presence of two giant planets embedded in the disc around HD\,163296. The analysis is performed using the  \discminer{} channel-map modelling framework on \twCOfull{} DSHARP data. Not only orbital radius, but also azimuth of the planets are retrieved by our technique. One of the candidate planets, detected at $R=94\pm6$\,au, $\phi=50\pm3^\circ$ (P94), is near the centre of one of the gaps in dust continuum emission, and is consistent with a planet mass of 1\,$\Mj$. The other planet, located at $R=261\pm4$\,au, $\phi=57\pm1^\circ$ (P261), is in the region where a velocity kink was previously observed in \twCO{} channel maps. Also, we provide a simultaneous description of the height and temperature of the upper and lower emitting surfaces of the disc, and propose the line width as a solid observable to track gas substructure. Using azimuthally averaged line width profiles we detect gas gaps at $R=38$\,au, $R=88$\,au, and $R=136$\,au, closely matching the location of their dust and kinematical counterparts. Furthermore, we observe strong azimuthal asymmetries in line widths around the gas gap at $R=88$\,au, possibly linked to turbulent motions driven by the P94 planet. Our results confirm that the \discminer{} is capable of finding localised, otherwise unseen velocity perturbations thanks to its robust statistical framework, but also that it is well suited for studies of the gas properties and vertical structure of protoplanetary discs.
\end{abstract}



\section{Introduction} \label{sec:intro}

Detecting planets in the early stages of formation is key to reconstructing the history and diversity of fully developed planetary systems, including our own. However, the dense and opaque environment where planets are assembled --protoplanetary discs-- make the direct observation of these bodies a challenging task. To date, PDS\,70 is the only system in which forming planets have been convincingly detected by direct imaging \citep{keppler+2018, haffert+2019}. Nevertheless, our growing understanding of how young planets interact with the disc material has stimulated the development of novel, albeit less direct, detection techniques. 

Embedded planets are expected to produce velocity disturbances observable in molecular line emission through the gaseous component of their hosting disc \citep{perez+2015, perez+2018}. In fact, there has been an increasing number of ALMA observations reporting localised \citep{pinte+2018b, casassus+2019, pinte+2019} and extended deviations from Keplerian rotation \citep{teague+2018a, teague+2019nat} attributed to the presence of planets, which have naturally inspired theoretical efforts on the characterisation and detection of planet-driven perturbations \citep{diskdynamics+2020, bollati+2021, rabago+2021, izquierdo+2021}. In particular, the disc around \hd{}, a Herbig Ae star at 101.5 pc from Earth \citep{bailer+2018}, has become one of the most interesting laboratories for the study of planet-disc interactions.  
It displays strong indications of embedded planets such as gaps and rings in the dust, and non-Keplerian deviations to the velocity field in the gas. 

As reported by \citet{isella+2016, isella+2018} using Band 6 continuum data, the dust gaps of the \hd{} disc are located at a radial distance of 10, 45, 86, and 141 au from the star (referred to as D10, D45, D86, D141), and the dust emission rings at 14, 67, 100, and 159 au (referred to as B14, B67, B100, B159). An additional dust gap appears at 270 au\footnote{Rescaled to the latest distance to the source.} (or D270) according to former observations in optical scattered light by \citet{grady+2000}. One way to explain these gaps is by invoking multiple embedded planets with masses between $\sim\!0.1-4\,\Mj$ depending on the physical properties of the disc \citep{zhang+2018}. However, multiple gaps driven by spiral waves from a single planet \citep{bae+2017}, or induced by non-planetary mechanisms should not be discarded \citep[see e.g.][for a review]{andrews+2020}.

Luckily, further constraints on the presence of planets have been possible thanks to recent kinematical analyses of the molecular gas in the disc. For instance, \citet{pinte+2018b} observed a localised velocity perturbation in \twCO{} channel maps, also known as a `kink', consistent with the presence of a 2\,$\Mj$ planet at an orbital distance of 260\,au according to hydrodynamic simulations. From here on we refer to this kink as K260. 
In the same disc, \citet{teague+2018a} detected azimuthally extended deviations from Keplerian rotation driven by radially localised pressure gradients typical of gas gaps \citep{kanagawa+2015}. By modelling the rotation curve of the system, the authors found a plausible scenario consisting of gas gaps carved by two Jupiter-mass planets orbiting at 83 and 137\,au. Moreover, at the radial location of these gaps, \citet{teague+2019nat} would later report the discovery of meridional circulation of gas flowing from the disc surface towards the midplane, providing further evidence for the presence of strong depletions in the surface density of the disc. 
However, these large-scale fluctuations associated with gas gaps should ideally be accompanied by the study of azimuthally localised perturbations in the velocity field in order to reduce the ambiguity that exists between planetary and non-planetary mechanisms to explain the origin of such substructures \citep[see e.g.][]{rabago+2021, izquierdo+2021}.

In this article, we apply the \discminer{} channel-map fitting analysis and statistical framework presented in \citet{izquierdo+2021} to search for embedded planets in the disc around \hd{} using DSHARP \twCOfull{} archival data. 
The technique detects two azimuthally localised velocity perturbations possibly driven by two giant planets; one at $R=94$\,au, $\phi=50^\circ$, within the D86 dust gap, and another at $R=261$\,au, $\phi=57^\circ$, potentially linked to the K260 velocity kink. Additionally, we use a best-fit model of the channel maps from the data to study the vertical structure of the disc, as well as the radial gradient of temperatures and line widths observed on the upper and lower emitting surfaces of \twCO{}.

While this paper was under review, further data on the source was published by the MAPS collaboration \citep{oberg+2021}. We have repeated the analysis of this work on the new data without finding differences affecting our conclusions. We will show a detailed analysis of the new data in a future publication.

\section{Line intensity model of the HD\,163296 disc}

\subsection{Dataset}

In this work we use \twCOfull{} line observations of the disc around \hd{} obtained by the DSHARP ALMA Large Program \citep{andrews+2018, isella+2016, isella+2018}. The synthesized beam of the data is $0\farcs{104}\times0\farcs{095}$, and the velocity channels are spaced by $0.32$\,km\,s$^{-1}$. The rms noise per channel is 0.84\,mJy\,beam$^{-1}$. The data cube is available at \url{https://almascience.eso.org/almadata/lp/DSHARP}. Details on the calibration of the data and imaging process can be found in \citet{andrews+2018}.

\subsection{Discminer model setup} \label{sec:discminer}

To model the line intensity and kinematics of the disc we use the \discminer{} package introduced in \citet[][hereafter Paper I]{izquierdo+2021}. The \discminer{} assumes parametric prescriptions for the peak intensity, line width, rotation velocity, and height of the emitting surfaces of the disc to produce intensity channel maps. It then invokes the \emcee{} 
sampler \citep{foreman+2013} to find the model parameters that best reproduce the intensity of the input data cube. This approach guarantees that the physical and morphological attributes of the disc are modelled simultaneously, providing a comprehensive picture of the gas disc structure and kinematics. 

The \discminer{} pieces the disc attributes together in a predefined kernel to generate model line profiles and channel maps. As in \citetalias{izquierdo+2021}, we adopt a generalised bell kernel to shape the model intensity, $I_m$, as a function of the disc cylindrical coordinates $(R,z)$ as follows,
\begin{equation} \label{eq:kernel}
    I_m(R, z; \upsilon_{\rm ch}) = I_p \left(1+\left|\frac{\upsilon_{\rm ch} - \upsilon_{\rm k^{l.o.s}}}{L_w} \right|^{2L_s} \right)^{-1},
\end{equation}
where $I_p$ is the peak intensity, $L_w$ is half the line width at half power, or just line width from now on, $L_s$ is the line slope. For simplicity, we parameterise these attributes as power laws of the disc cylindrical coordinates ($R$, $z$). On the other hand, $\upsilon_{\rm ch}$ is the channel velocity where the intensity is to be computed, and $\upsilon_{\rm k^{l.o.s}}$ is the Keplerian line-of-sight velocity. 
The vertical coordinate $z$ is determined by the height of the upper and lower emission surfaces, which implies that each attribute (but the line slope) has two different representations. The height of each surface is parameterised independently using a combination of two radial power laws. The exact functional form of each attribute and the free parameters of the model are summarised in Table \ref{table:attributes_parameters}.

To merge the contribution of the upper and lower emitting surfaces into a single line profile, on each velocity channel and pixel we select the highest intensity between bell profiles computed for both surfaces independently. In terms of radiative transfer this type of masking is more precise than adding up both intensity profiles directly. The reason is that in a real scenario, the emission from the lower surface can only be distinguished when the upper surface becomes optically thin enough, which for \twCO{} is mainly limited to the wings of the profile\footnote{Assuming that the sensitivity, and the angular and spectral resolution of the observation are sufficient to resolve both emitting surfaces.}.

It should be noted that the model attributes introduced here are merely descriptive and are not the result of detailed radiative transfer. Also, these are constrained to the upper and lower emitting surfaces of \twCOfull{}, and therefore any extrapolation to other scale heights in the disc should be done with caution.

\subsection{Parameter search with \emcee{}}
\label{subsec:emcee}

%-----------------------------------------------------------------
\begin{figure*}
   \centering
   \includegraphics[width=1.0\textwidth]{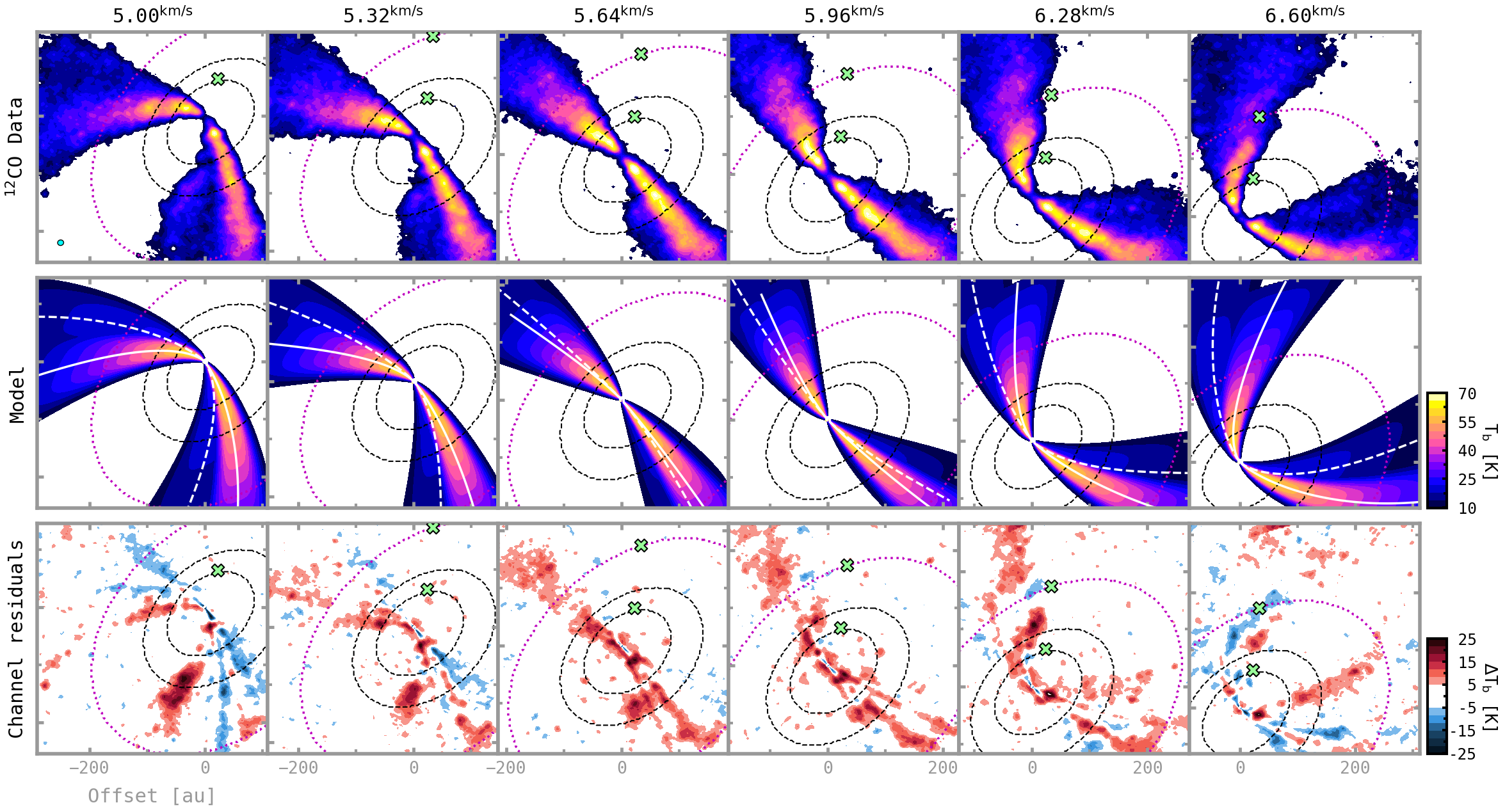} 
      \caption{ Selected channel maps from DSHARP \twCOfull{} data of the \hd{} disc (top row), compared to those from the best-fit model found by the \discminer{} (middle row). Line-of-sight velocity contours from the model upper and lower surfaces of the disc are overlaid on the model channel maps as solid and dashed white lines, respectively. Also shown are residual brightness temperatures for each velocity channel (bottom row). For reference, the best-fit systemic velocity is $\upsilon_{\rm sys}=5.77$\,km\,s$^{-1}$. The synthesized beam of the observation is shown in the bottom left of the top left panel as a cyan ellipse. Velocity channels are spaced by 0.32\,km\,s$^{-1}$. Residuals with magnitudes lower than the rms noise of the data were masked. Green crosses highlight the position of the localised velocity perturbations, P94 and P261, detected by our clustering algorithm in Sect. \ref{subsec:planets}. Dashed black lines mark the projected location of the D86 and D141 dust gaps \citep{isella+2018}, while the dotted purple line shows the radial location of the K260 kink \citep{pinte+2018b}. From visual inspection, a localised kink-like feature near the P94 perturbation is observed in the $\upsilon_{\rm ch}=5.96,\,6.28$\,km\,s$^{-1}$ channels. The K260 kink near the P261 perturbation is more extended, spanning from $\upsilon_{\rm ch}=5.96$ to at least $\upsilon_{\rm ch}=6.92$\,km\,s$^{-1}$ (see also Fig. \ref{fig:kink_channels}).  
              }
         \label{fig:channels}
\end{figure*} 

%-----------------------------------------------------------------

We initialise the \emcee{} sampler with a first guess of parameters according to previous measurements of the inclination, position angle, and stellar mass \citep{isella+2018,teague+2019nat}. The other initial parameters are guessed using the prototyping tool of the \discminer{} which allows for a quick comparison of the morphology of model channel maps with respect to the data. The MCMC search is performed with 256 walkers which evolve for 3000 steps for an initial burn-in stage. Next, to sample the posterior distributions and assess convergence of parameters we run the same number of walkers for 10000, 20000, and 50000 steps. We note that the variance and the median of parameter walkers remain almost unchanged after $\sim\!10000$ steps.
The best-fit parameters summarised in Table \ref{table:attributes_parameters} are the median of the posterior distributions in the last 5000 steps of the 20000 step run, thinned by half the auto-correlation times of the parameter chains in order to minimise the impact of non-independent samples on the posterior statistics. In Figure \ref{fig:channels}, we compare selected channel maps from the data and the best-fit model obtained with these parameters. 

On a side note, we observe that the best-fit stellar mass retrieved by our model is affected by the choice of the disc outer radius. For a disc radius of $R_d=380$\,au we find a stellar mass of $M_\star=2.02$\,M$_\odot$, 
whereas for $R_d=450$\,au the stellar mass decreases to $M_\star=1.97$\,M$_\odot$. This behaviour is expected as the model tries to compensate for the sub-Keplerian rotation supported by steep pressure gradients at large disc radii \citep[][]{dullemond+2020}. However, none of the results of this paper is affected by such a small variation in the stellar mass.

The noise of the data is taken into consideration for the parameter search as follows. At each pixel of the data, we compute the standard deviation of the residual intensities in line-free channels and take it as the weighting factor of the likelihood function to be maximised by the sampler \citepalias[see Eq. 1 of][]{izquierdo+2021}. To ensure that the noise of individual pixels is approximately independent from neighbouring pixels, we down-sample the data and the model grid so that the pixels are separated by $\sim$1.5 beams between each other. Also, this enables us to safely consider the variance of the posterior distribution of model parameters and their (anti--)correlations to quantify uncertainties in observables derived from the model. Details on the analytic propagation of errors taking into account parameter correlations are presented in Appendix \ref{sec:appendix_errors}.

\section{Results} \label{sec:results}

\subsection{Physical attributes of the \twCO{} disc} 

%-----------------------------------------------------------------
\begin{figure*}
   \centering
   \includegraphics[width=1.0\textwidth]{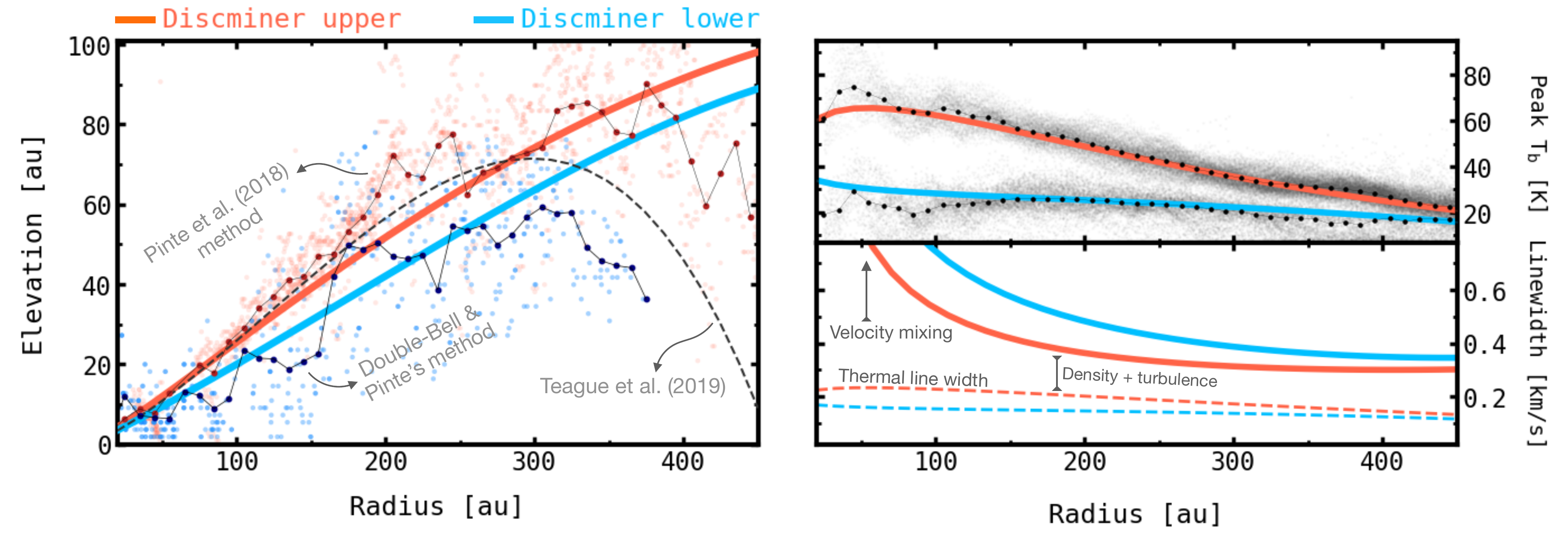}
      \caption{Best-fit attributes derived by the \discminer{} for the upper (solid orange) and lower (solid blue) emitting surfaces of \twCOfull{} in the disc of HD\,163296. The functional forms of the model attributes and best-fit parameters are presented in Table \ref{table:attributes_parameters}. \textit{Left}: Height of the model emission surfaces. The background dots illustrate the height of the upper (orange) and lower (blue) surfaces derived independently with the method of \citet{pinte+2018a}. See Sect. \ref{subsec:height} for details on how to use the method on the lower surface alone. The connected circles mark the mean value of dots within 10\,au ($\sim$beam size) intervals. Also shown is the height reported by \citet{teague+2019nat}. \textit{Top right}: Model peak brightness temperature, computed with the full Planck law. The background dots are the primary and secondary peaks, associated with emission from the upper and lower surface, respectively, as retrieved from double-Bell profiles fitted along the velocity axis of the data cube (see Sec \ref{subsec:height}). \textit{Bottom right}: Model half line widths at half maximum. The dashed lines indicate half the thermal broadening at half maximum for \twCO{} at the model temperatures of the top right panel. The difference between the thermal and the observed supra-thermal line width is given by the contribution of density and turbulence to the optical depth of the line and by velocity mixing. 
              }
         \label{fig:emission_surfaces}
\end{figure*}
%-----------------------------------------------------------------
In this Section, we discuss the form of the main physical attributes retrieved by the \discminer{} model of the \twCOfull{} emission from the disc around \hd{}. In Figure \ref{fig:emission_surfaces}, peak brightness temperature, line width and height of the upper and lower emitting surfaces are shown as a function of the radial location in the disc.

\subsubsection{Emission height}
\label{subsec:height}

Our best-fit model indicates that the scale height of the upper surface of the disc is around $z/R\approx0.26$, similar to the findings of \citet{rosenfeld+2013}. This scale height is also in good agreement with the kinematical model of the upper surface reported by \citet{teague+2019nat}, although they diverge substantially on the outskirts of the disc ($R>300$\,au). For further comparison, we also determined the height of the upper surface using the \textsc{disksurf} code \citep{teague+2021disksurf}, which is an open source implementation of the geometrical method introduced by \citet{pinte+2018a} for measuring the altitude of molecular line emission in discs. This independent experiment is better reproduced by the upper surface of the \discminer{} up to $R=400$\, au, although beyond that radius the three methods seem to differ. One of the reasons for this discrepancy may be the fact that the boundary to distinguish intensities from the upper and lower surfaces is diffuse towards the edge of the disc, and hence the extraction of the upper surface might be biased by the contribution of the lower counterpart. Nevertheless, our analysis of gas substructures and detection of planets takes place within $R<300$\,au, where the three measurements agree.

Unlike previous methods, the \discminer{} allows us to infer the height of the lower emitting surface too. Our model finds that the lower surface stands around $z/R\approx0.2$ scale heights above the midplane of the disc. To validate this part of the modelling, we performed an empirical reconstruction of the lower surface intensity from the data and then estimated the altitude of its emission using once again the geometrical method of \citet{pinte+2018a}. To do this, we first fit double-bell profiles along the velocity axis of the pixels of the data cube in an attempt to separate the upper and lower surface emission. Next, on each pixel we extract the bell profile with the smaller peak intensity of the two, which is normally associated with the lower surface contribution to the line intensity profile \citep[see e.g.][]{dullemond+2020}. Finally, we combine these secondary bell profiles from all pixels to generate channel maps of the lower surface alone. These channel maps can then be an input to the method of \citet{pinte+2018a} to determine the altitude of the lower surface emission, which is illustrated by the blue dots in Fig. \ref{fig:emission_surfaces}, left panel. The \discminer{} height of the lower surface is consistent with this independent estimate, at least for $R<300$\,au where our following analyses take place. Centroid velocities of the lower and upper surfaces obtained separately from this empirical reconstruction are presented in Fig. \ref{fig:centroids_low_up}.

\subsubsection{Line width and Brightness temperature}
\label{subsec:linewidth}

Because of the high densities present in protoplanetary discs, the emission from abundant molecules such as \twCO{} is optically thick almost everywhere. For a related reason, the level populations of these molecules can be safely considered in local thermodynamic equilibrium \citep{weaver+2018}. Both these facts imply that the peak intensity at the central channel of the line emission converges to the kinetic temperature of the gas, and it saturates over neighbouring channels until the optical depth becomes low at the line wings. Thus, the extent of the plateau at the top of the line is highly influenced by the density of the species, and consequently so is the observed line width. Conversely, if the transition was optically thin, the line broadening would be primarily dominated by thermal and turbulent motions \citep{hacar+2016}.

While the thermal broadening at half maximum for \twCO{} at 50\,K is 0.21\,km\,s$^{-1}$, the best-fit model line width at the same temperature (i.e. at $R=140$\,au, on the upper surface) can be as high as 0.43\,km\,s$^{-1}$, and the contrast becomes even larger on the lower surface of the disc (0.15 vs 0.60\,km\,s$^{-1}$). This noticeable excess in line widths across the entire disc could be explained by strong turbulent motions in the gas.  
However, using observations of the \hd{} disc with different molecules, \citet{flaherty+2017} obtained upper limits for turbulent broadening of only $\delta_{\rm turb}<0.06c_s$, where $c_s$ is the sound speed of the medium ($c_s=0.50$\,km\,s$^{-1}$ at 50\,K, assuming a unit mass for the medium of $\mu=2.37$\,u), which means that the supra-thermal line widths should instead be dominated by the density of the species in most of the disc\footnote{At small radii, due to the finite angular and spectral resolutions, velocity mixing becomes the dominant source of line broadening.}. Such an opacity effect could explain why the retrieved line widths on the lower surface are generally higher than those on the upper surface despite the fact that the lower surface is considerably cooler. The width of the \twCO{} line profile is thus an indirect window to the gas density, which we exploit in Sect. \ref{subsec:gaps} to determine the location of gas gaps and analyse asymmetries in the gas substructure. 

The peak brightness temperature of the model upper surface is consistent with the peak intensities measured directly from the data, which span a wide range of temperatures between $\sim20\!-\!70$\,K \citep[see also][]{isella+2018}. The peak brightness temperatures of the model lower surface oscillate between $\sim20\!-\!30$\,K, which is compatible with temperatures where CO molecules are expected to freeze-out onto dust grains \citep[see e.g.,][]{miotello+2014, woitke+2016}. They also agree with previous radiative transfer models \citep{flaherty+2015}, and direct estimates \citep{dullemond+2020}, as well as with the secondary peak intensities retrieved independently by the double-bell fits introduced in Sect. \ref{subsec:height}.

\subsection{Residual maps} \label{subsec:residuals}

%-----------------------------------------------------------------
\begin{figure*}
   \centering
   \includegraphics[width=1.0\textwidth]{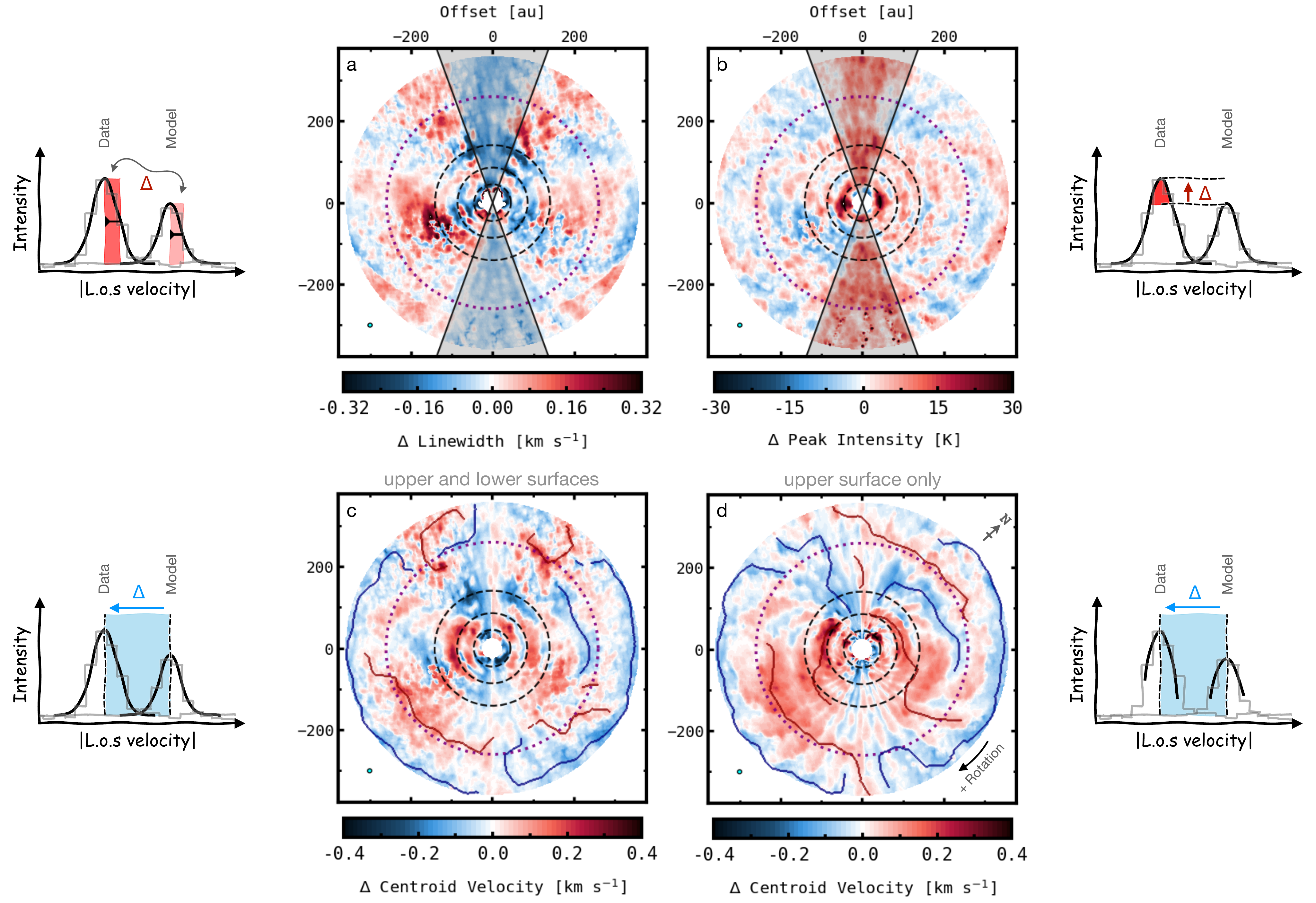}
      \caption{Deprojected residuals retrieved by the \discminer{} (a, b, c) and \textsc{bettermoments} (d) by comparing \twCOfull{} observations of the disc around \hd{} and best-fit line profiles found by the \discminer{} as described in Sect. \ref{subsec:residuals}. The cartoons next to each panel illustrate how the corresponding residuals are computed. The synthesized beam of the observation is shown in the bottom left of all panels as a cyan ellipse. The beam is displayed as it is projected on the sky plane. For reference, the north axis of the sky plane and the rotation direction of the disc are marked in panel d. \textit{Top row}: (a) Line width and (b) peak intensity residuals. The shadows indicate regions excluded from the analysis due to systematic residuals possibly caused by velocity averaging in channels around the projected minor axis of the disc. \textit{Bottom row}: Velocity residuals obtained by subtracting the absolute value of data and model line centroids, which are (c) computed from full line profiles to account for the contribution of the upper and lower surfaces of the disc, or (d) measured around the peak of line profiles to which the upper surface contributes the most. Blue and red lines overlaid on velocity residuals are the spines of sub-- and super-Keplerian filamentary structures found by \textsc{filfinder}. The dashed lines indicate the location of the D45, D86 and D141 dust continuum gaps registered by \citet{isella+2018}. The outer dotted line marks the radial distance of the K260 kink reported by \citet{pinte+2018b}.  
              }
         \label{fig:residuals_2d}
\end{figure*}
%-----------------------------------------------------------------

To extract observables and quantify line profile differences between the data cube and the smooth Keplerian model of the disc, we fit a Gaussian profile to each pixel of the data and the model cube\footnote{We fit Gaussian and not double-bell profiles because the data line centroids from the latter have high pixel-to-pixel variations, of the order of the channel width. This stage is not to be confused with the generation of model line profiles for the MCMC minimisation of intensity differences discussed in Sect. \ref{sec:discminer}.}. 
The standard deviation, amplitude, and mean value of each Gaussian profile represent, respectively, the line width, peak intensity, and centroid velocity of the corresponding pixel. The Gaussian properties of the model line profiles are then subtracted from those of the data to produce line width, peak intensity and centroid velocity residuals, as illustrated in Figure \ref{fig:residuals_2d}. Analogous to \citetalias{izquierdo+2021}, the velocity residuals reported in this work are defined as the difference between the absolute value of data line centroids and the absolute value of model line centroids. Considering the absolute value before subtraction of velocities can be convenient for visualisation because it makes residuals on the blueshifted side of the disc to switch sign with respect to residuals computed from direct subtraction. This occurs in such a way that sub(super) Keplerian perturbations in the azimuthal component of the velocity, as those expected around gas gaps, appear blue(red) in the residuals map. Differences between velocity residuals computed by direct subtraction and by subtraction of absolute values of line centroids are illustrated in Appendix \ref{sec:supporting_figures}, Fig. \ref{fig:kind_residuals}.

It should be noted that our forward-modelling of channel maps allows us to account for the effect of intensity variations on the retrieval of gas velocities from both model and data. For this reason, fitting Gaussians to full line profiles can be safely done. Note that this approach implies that the retrieved centroid velocities and line widths are the result of the combined contribution of the upper and lower surfaces to the intensity of the disc. For comparison, we explore another possibility using the quadratic fit method supported by the \textsc{bettermoments} package \citep{teague+2018_bettermoments} which operates with intensity channels around line profile peaks to determine line-of-sight velocities, primarily representative of the upper surface of the disc (see Fig. \ref{fig:residuals_2d}d).  

Additionally, we use both kinds of velocity residuals, `upper+lower' and `upper--only', to find coherent filamentary structures with the \textsc{filfinder} package \citep{koch+2015}. To start the search, we assume a smoothing size of 10\,au, similar to the extent of the synthesized beam of the data, and a minimum size of 1500 pixels for a filament to be considered in the analysis. The red and blue lines overplotted in the bottom panels of Fig. \ref{fig:residuals_2d} are the medial axes of the filamentary structures found by the algorithm in our velocity residuals. In contrast, intensity and line width residuals do not exhibit elongated signatures as clearly. Nevertheless, in Sect. \ref{subsec:gaps} we show that all three types of residuals provide remarkable clues about the gas substructure in the disc. 

On the other hand, velocity residuals can also be used to hunt for candidate embedded planets. As demonstrated in \citetalias{izquierdo+2021}, the presence of a planet is closely related to spatially localised velocity perturbations in the gas disc, whose magnitude and location should be retrievable as long as the resolution and signal-to-noise ratio of the data allow it.

In what remains, we focus our analysis on these three types of residuals to track gas substructure, and to search for localised velocity perturbations possibly associated with the presence of young planets in the disc of \hd{}. 

\subsection{Gas gaps} \label{subsec:gaps}

%-----------------------------------------------------------------
\begin{figure*}
   \centering
   \includegraphics[width=0.95\textwidth]{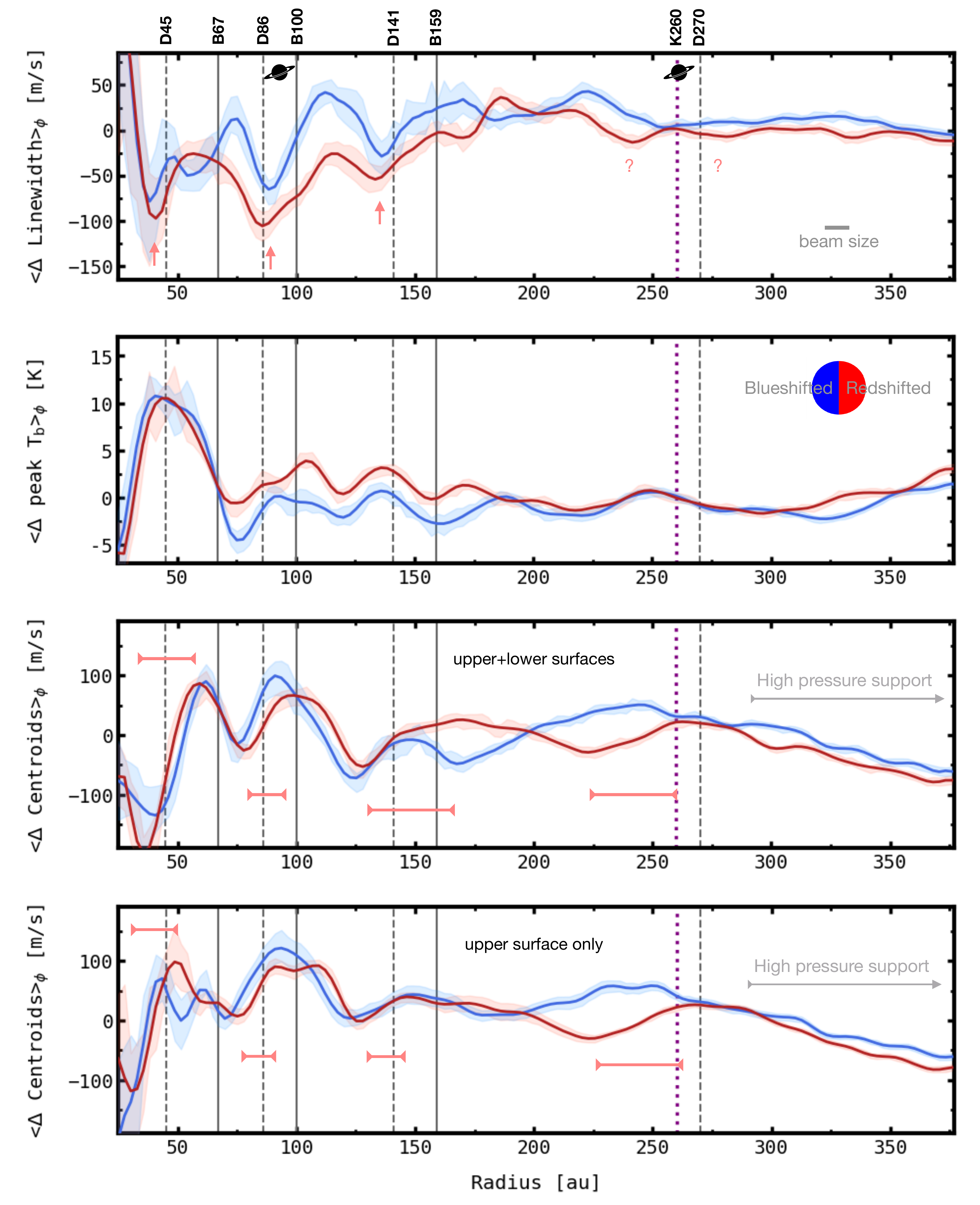} 
      \caption{Azimuthally averaged line width, peak intensity and centroid velocity residuals extracted from the blue-- and redshifted halves of the disc, computed within $\pm70^\circ$ azimuths to avoid systematic residuals near the disc minor axis (see Fig. \ref{fig:residuals_2d}, top row). The radial location of dust gaps, dust rings, and the K260 kink is marked by the dashed, solid and dotted lines, respectively. In the top panel is also shown the radial location of the two planet detections reported in Sect. \ref{subsec:planets}. The shaded regions represent one standard deviation from the mean value, divided by the square root of the number of independent beams along each projected annulus. Gas gaps on the redshifted side of the disc are indicated by arrows and brackets; arrows mark local line width minima near dust gaps, while brackets enclose positive velocity gradients indicative of localised variations in the gas pressure.
              }
         \label{fig:azimuthal_average}
\end{figure*} 
%-----------------------------------------------------------------

As discussed in Sect. \ref{subsec:linewidth}, the line broadening of highly optically thick tracers is dominated by the species density, which in other words means that only a fraction of the line width originates from (non--)thermal motions \citep{hacar+2016}. Therefore, \twCO{}-depleted regions in discs are expected to drive prominent line width minima as illustrated in \citetalias{izquierdo+2021} for a planet-carved gap.
Line centroids are also sensitive to the presence of gaps due to the fact that any gas substructure triggers local pressure forces which induce deviations from Keplerian rotation that follow the geometry of the pressure gradient. In the gas gap scenario, one can expect axisymmetric velocity perturbations with a positive radial gradient enclosed within the edges of the gap \citep[][]{kanagawa+2015, teague+2018a}. To exploit this background knowledge and search for gas gaps in the disc of \hd{}, we compute azimuthally averaged profiles of line width, peak intensity, and velocity residuals as displayed in Figure \ref{fig:azimuthal_average}. This is done separately for the red-- and blueshifted halves of the disc to highlight large-scale azimuthal asymmetries.

\subsubsection{Azimuthally averaged residuals}

From the average line width and velocity residual profiles there are clear indications of gas gaps near the D45, D86 and D141 dust gaps. Line width residuals exhibit local minima at $R=38$\,au, $R=88$\,au and $R=136$\,au, closely coexisting with positive velocity gradients, and in good agreement with the location of gas gaps proposed by the radiative transfer modelling of \citet{isella+2016} and \citet{liu+2018}. Also, the positive velocity gradients observed in our average velocity residuals are consistent with rotation curves of the same disc reported in recent studies \citep{teague+2018a, rosotti+2020}. We note additional evidence of gas substructure near the K260 kink and the D270 dust gap, although this time the line width trough is not as clear as the closest positive velocity gradient, centred around $R=245$\,au on the redshifted half of the disc for both types of centroid residuals.

On the other hand, the average peak brightness temperature displays several local minima that do not overlap with the line width minima, nor with dust gaps or rings. Instead, some local maxima seem to co-locate with line width and dust gaps (D45\footnote{However, the temperature peak at D45 is likely due to continuum subtraction and beam dilution in the inner 30\,au of the disc \citep{teague+2019nat}. This behaviour is captured by the model at the cost of underestimating temperatures around 50\,au (see also Fig. \ref{fig:emission_surfaces}, top right).}, D141), but there are multiple exceptions. Thus, looking at line width fluctuations in optically thick tracers appears to be a more reliable alternative to probing gas substructure. The reason is presumably that the line width is measured over the full line profile, so it can trace the systematic effect that the varying optical depth spawns over different velocity channels, while the peak intensity is measured on a single channel, making it more subject to local thermodynamic fluctuations and noise. Moreover, this means that the way in which gas gap attributes are retrieved could strongly impact the interpretation of hydrodynamic properties of discs and, more specifically, the inference of planet masses from radiative transfer models of gaps.

\subsubsection{Non-axisymmetric gas substructure}

Our separate analysis of both halves of the disc allows us to comment on asymmetries present in the gas distribution and kinematics. Asymmetries in the gas velocities are subtle throughout most of the disc, but near the location of the K260 kink there is a significant radial shift of about $\sim\!50$\,au between the peak of positive velocity gradients on each half of the disc. This is possibly an effect of non-axisymmetric spiral-like perturbations (see Fig. \ref{fig:residuals_2d}, bottom row), potentially linked to the hydrodynamic interaction of the disc and the massive planet detected by \citet{pinte+2018b} at $R=260$\,au, on the redshifted side. Another substantial asymmetry in average gas velocities is observed between 140\,au and 200\,au. However, this feature appears on the `upper+lower' centroid residuals only, suggesting that instead of an actual asymmetry in the velocity field it could be related to contamination of the lower surface emission by dust absorption in the midplane of the disc \citep[see][]{isella+2018}. 

Unsurprisingly, there are asymmetries in the azimuthally averaged line width and intensity profiles too. Line widths are systematically lower on the redshifted part of the disc, between about B67 and B159, while peak intensities are slightly higher than those on the blueshifted side, in the same radial section. This finding reveals azimuthal fluctuations in the density and temperature of the disc: the blueshifted side of the disc is denser and cooler than the redshifted half. A plausible origin of these asymmetries may be the presence of massive planets, which are capable of producing vertical and turbulent motions that induce azimuthal gradients of velocity dispersion, density and temperature around their orbit \citep{dong+2019}.
The spatial coincidence of these features with the D86 and D141 gaps is in agreement with such a scenario. In particular, the blueshifted side of the disc has a prominent line width excess of $\sim\!50$\,m\,s$^{-1}$ in the D86 gap, which is similar to the expected excess that a massive planet would trigger in and around its gap \citep[see Fig. 8 of][]{dong+2019}. In Sect. \ref{subsec:planets}, we detect an azimuthally localised velocity perturbation near D86 which strengthens the idea of an embedded planet at this location. Future studies of optically thin isotopologues could offer additional clues about candidate massive planets by characterising line width and temperature asymmetries in the disc.

\subsection{Detection of planets} \label{subsec:planets} 

%-----------------------------------------------------------------
\begin{figure*}
   \centering
   \includegraphics[width=0.83\textwidth]{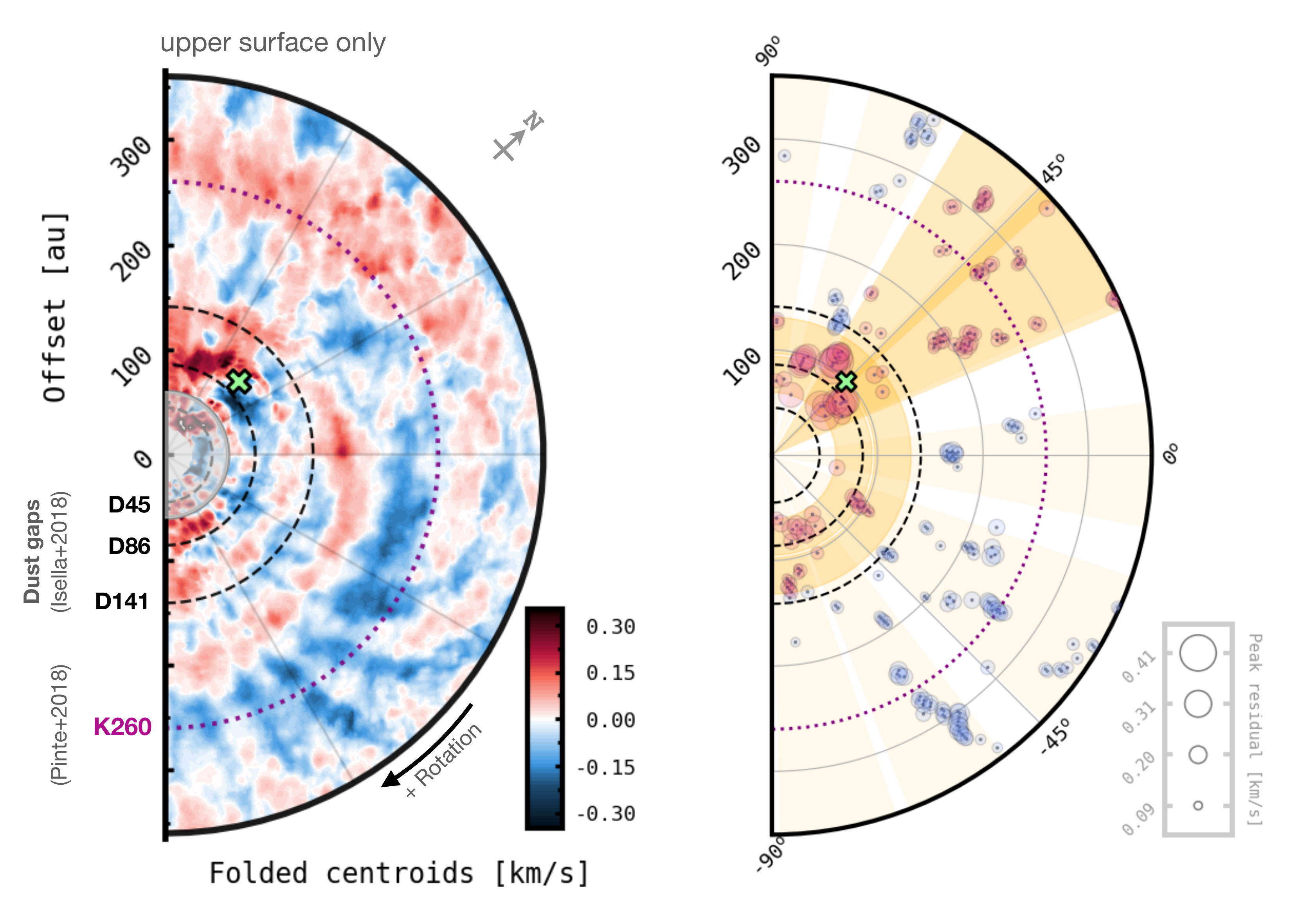}\\ \vspace{0cm}
   \includegraphics[width=0.83\textwidth]{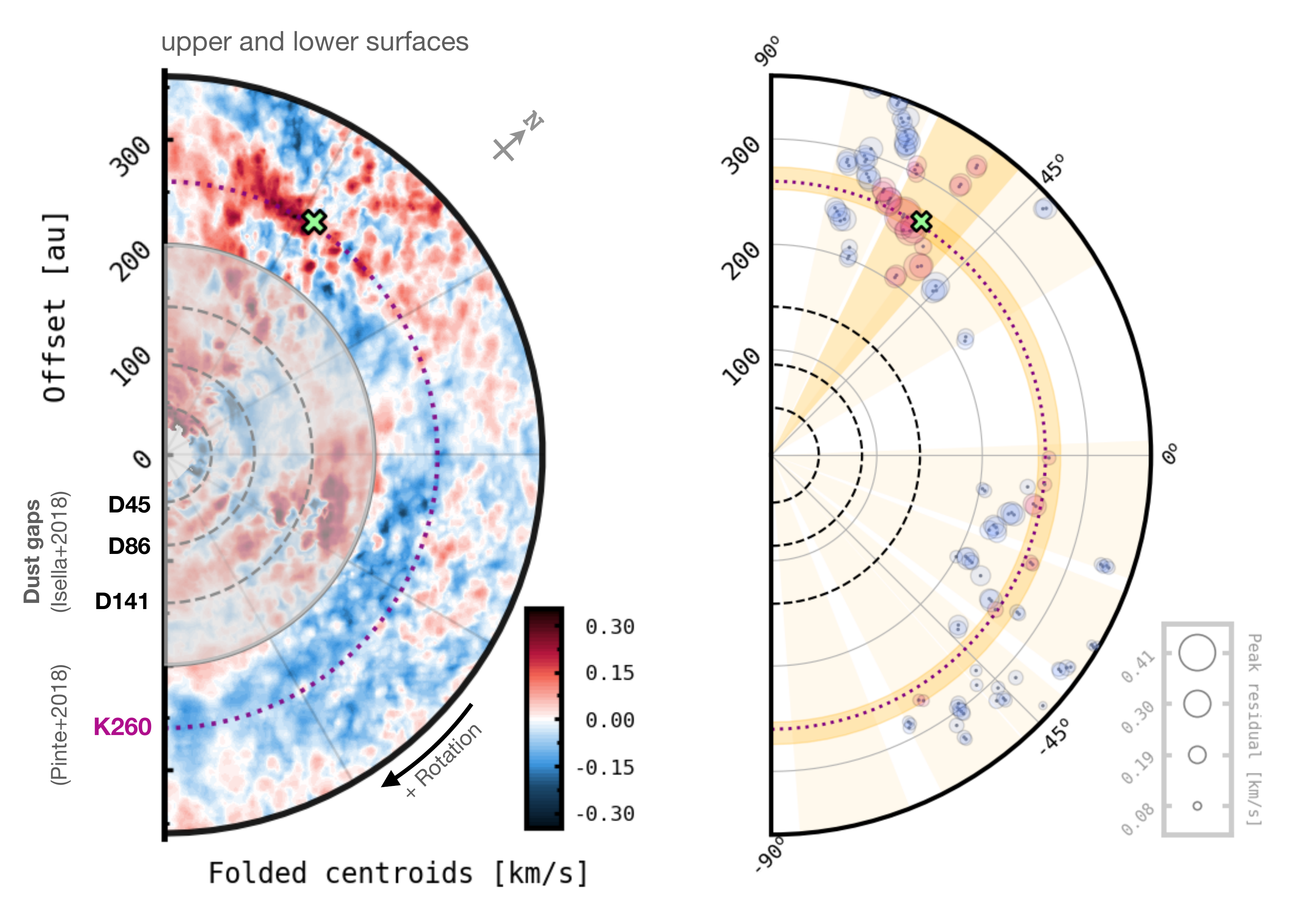}
  
      \caption{Folded velocity residuals (left column) and detected clusters of peak velocities (right) using seven radial and eight azimuthal clusters. The centroid velocities are measured either around the peak of the line profiles, namely on the upper surface of the disc (top row), or using full line profiles, that is, considering the combined contribution of the upper and lower surfaces (bottom row). The green crosses mark the inferred position of the localised velocity perturbations detected by our clustering algorithm, P94 and P261, possibly driven by embedded planets. The grey regions in the left panels correspond to masked portions due to high velocity uncertainties (top; see Appendix \ref{sec:appendix_errors}), or possible contamination of the lower surface by dust absorption in the midplane (bottom).
              }
         \label{fig:planet_detection}
\end{figure*} 
%-----------------------------------------------------------------

We employ the statistical analysis developed in \citetalias{izquierdo+2021} to search for localised planet-driven perturbations in the gas disc kinematics of \hd{}. More specifically, we use the so-called Variance Peak method of that framework to exploit the fact that peak deviations from Keplerian rotation are expected to be spatially clustered near planets. 

The first step of the Variance Peak method consists of folding centroid velocity residuals (Fig. \ref{fig:residuals_2d}, bottom row) along the projected minor axis of the disc to get rid of any axisymmetric feature driven by symmetric substructures such as gas gaps. 
In other words, we subtract line centroids on the blueshifted side of the disc from their mirror location on the redshifted half. Next, a radial scan is performed over the map of folded centroids to search for peak velocity residuals, and to record their magnitude, their azimuth, and radial location, as displayed in Figure \ref{fig:folded_azimuthal}. With this information we run a K-means clustering algorithm \citep{macqueen+1967, scikit-learn+2011} along the radial and azimuthal coordinates, independently, to look for localised velocity perturbations. 
The K-means algorithm subdivides the input residuals into a predefined number of clusters in such a way that the centre of each cluster is the closest centre to all the residuals in the cluster. Said differently, the input data are iteratively partitioned into Voronoi cells until convergence is reached, which in this case means until the sum of squared distances from the peak residuals to the centre of their clusters is minimised. We refer the reader to \citetalias{izquierdo+2021}, Sect. 4.4, for further details on the K-means algorithm applied to the analysis of velocity residuals.
If a velocity perturbation is strong and coherent, the variance of its corresponding cluster of peak velocity residuals should be high and exceed the variance of the background clusters, as predicted in \citetalias{izquierdo+2021}. If such an excess is higher than three times the dispersion of the background variances (i.e. $>\!3\sigma$) we claim for the presence of a localised perturbation possibly driven by a planet.

This is illustrated in Figure \ref{fig:planet_detection}, where we present folded centroid velocity maps for one half of the deprojected disc, and the localised perturbations detected by the Variance Peak method. The detection algorithm is applied on centroid velocities derived from the upper surface alone (top row), and from the combined contribution of the upper and lower surfaces (bottom row). For the latter case, we masked the inner part of the disc where there is obscuration of the lower surface emission due to dust absorption in the midplane \citep{isella+2018}. Non-uniform absorption features would trigger spurious velocity fluctuations that may bias the detection process.

\subsubsection{P94 perturbation}
\label{sec:p94}

On the upper surface, our method finds a strong localised perturbation with an amplitude of $\delta\upsilon=0.41$\,km\,s$^{-1}$, centred at $R=94\pm6$\,au, $\phi=50\pm3^\circ$ (hereafter P94)\footnote{Because of the folding procedure applied to the velocity residuals, the azimuth of the detected perturbation is degenerate with respect to the other half of the disc. That is, the same perturbation is found, with opposite sign, at $\phi=180^\circ-50^\circ=123^\circ$, near a candidate kink reported by \citet{pinte+2020} in the D86 gap. Nevertheless, since the localised perturbation has a `Doppler flip' morphology, we favour the detection at $\phi=50^\circ$ where the super-Keplerian part of it is exterior to the orbit of the planet, and the sub-Keplerian part is interior. This is expected so long the observed perturbation is not strongly dominated by radial velocity fluctuations, which is true for massive planets at azimuths between $\phi=0^\circ$ and $\phi=70^\circ$, as illustrated in Fig. 5 of \citetalias{izquierdo+2021}.} in the disc frame of reference. This perturbation is in good agreement with the presence of a planet previously hypothesized as the main driver of the D86 dust gap observed in continuum \citep{isella+2018}, and of the corresponding gas gap inferred from pressure bumps in the rotation curve of the disc \citep{teague+2018a}. The prominent line width asymmetries around the D86 gap further support the presence of this planet as discussed in Sect. \ref{subsec:gaps}. Also, P94 features a Doppler flip typical of spiral wakes generated by an embedded planet \citep{perez+2018, casassus+2019}. We provide a rough estimate of the mass of this planet by rescaling the hydrodynamic simulations presented in \citetalias{izquierdo+2021}, for a stellar mass of 2\,M$_{\odot}$ and a planet at $R=94$\,au (see Fig. \ref{fig:simulations}). Omitting radiative transfer effects, our simulations would predict that a 3\,$\Mj$ planet can produce perturbations with an amplitude similar to that of P94. 
However, in \citetalias{izquierdo+2021} we demonstrated that peak velocity fluctuations observed in folded velocity residuals can be amplified due to radiative transfer and projection effects. In particular, in Figs. 5 and 10 of that work we showed that at intermediate azimuths, between $\phi\approx30^\circ$ and $\phi\approx60^\circ$, an intrinsic perturbation of $\sim0.25$\,km\,s$^{-1}$ can become as large as $\sim0.4$\,km\,s$^{-1}$ when both effects are considered. Extending that result to this scenario, and by inspection of Fig. \ref{fig:simulations}, left panel, a 1\,$\Mj$ planet would be sufficient to explain the P94 perturbation, which is the same planet mass suggested by \citet{teague+2018a} near the D86 gap.
From \citetalias{izquierdo+2021}, we also note that line width residuals as low as the observed $\Delta L_w\approx-0.1$\,km\,s$^{-1}$ at this radius are compatible with a deep gas gap carved by a $1\,\Mj$ planet too.
Nevertheless, our mass estimate conflicts with the intermediate planet masses ($0.1\!-0.3\,\Mj$) proposed by \citet{zhang+2018} at D86 based on hydrodynamic simulations of the dust gaps in the disc. This tension is expected because the dynamical properties of dust, which dictate how the dust grains interact with the gas in the disc, are still poorly constrained by observations. On the contrary, the use of kinematic measurements of planetary masses can help model the local properties of the dust with greater precision \citep{pinte+2019}. 
Nevertheless, we note that the accuracy of planet mass estimates from forward modelling of hydro simulations can still be systematically affected by a simplistic treatment of the gas thermodynamics \citep[see e.g.][]{bae+2021}.

\subsubsection{P261 perturbation}
\label{sec:p261}

From the upper surface velocity residuals we do not find azimuthally localised perturbations linked to the K260 kink registered by \citet{pinte+2018b}. Instead, we report that such a kink is actually driven by a long filamentary structure in the gas kinematics, spanning from $\sim\!90^\circ$ to $20^\circ$ azimuths, and centred around $R\approx270$\,au. This is illustrated in Figure \ref{fig:kink_channels}, where we compare isovelocity contours of the data against those of the \discminer{} model, and highlight that the kink is present over the channels that cross the $\delta\upsilon\approx0.15$\,km\,s$^{-1}$ filament on the top right corner of the deprojected disc. Conversely, there are a few other elongated deviations from Keplerian rotation that do not exhibit clear kinks to the eye because they are either weaker or stand at azimuths close to the main axes of the disc.

Nevertheless, when the combined contribution of the upper and lower surfaces is considered, a localised fluctuation of magnitude $\delta\upsilon=0.40$\,km\,s$^{-1}$ appears at $R=261\pm4$\,au, $\phi=57\pm1^\circ$, in the region associated with the K260 kink. In \citetalias{izquierdo+2021}, we have shown that planet-driven perturbations are sometimes best observed on the lower emitting surface due to projection effects, especially for planets at intermediate azimuths in the near side of the disc, as in this case. Furthermore, it is also known that the magnitude of the three-dimensional velocity perturbations around a planet change with scale-height \citep[see e.g.,][]{rabago+2021}, which could explain why the morphology and magnitude of the perturbation vary when the lower surface of the disc is taken into account.  
Both effects strengthen the idea that this localised perturbation (hereafter P261) should be caused by an embedded planet, which at the same time is likely to be the main driver of the long kinematic filament associated with the K260 kink.

\subsubsection{Detection significance of the localised perturbations}

The significance of the P94 and P261 perturbations is above an acceptance threshold of $3\sigma$ with respect to the background velocity residuals. We note that both localised signatures are robustly detected regardless of the number of clusters considered for the K-Means algorithm, which we tested using six to ten clusters. The P94 detection yields an average significance of $19.4\sigma$ in radius and $7.5\sigma$ in azimuth, while the P261 detection has an average significance of $5.2\sigma$ in radius and $4.6\sigma$ in azimuth.
Typical mean values of background cluster variances are between (0.3--0.4)$\times 10^{-3}$ and (0.5--0.6)$\times 10^{-3}$\,km$^2$\,s$^{-2}$ for the P94 and P261 analyses, respectively, while 1$\sigma$ values are within (0.1--0.2)$\times 10^{-3}$ and (0.3--0.4)$\times 10^{-3}$\,km$^2$\,s$^{-2}$, where $\sigma$ represents the standard deviation of background cluster variances. The reported orbital radius and azimuth of P94 and P261 is the mean value of the detected location weighted by the statistical significance of the measurement, while the reported uncertainty is the weighted standard deviation of the detected locations in all realisations.

%-----------------------------------------------------------------
\begin{figure*}
   \centering
   \includegraphics[width=1.0\textwidth]{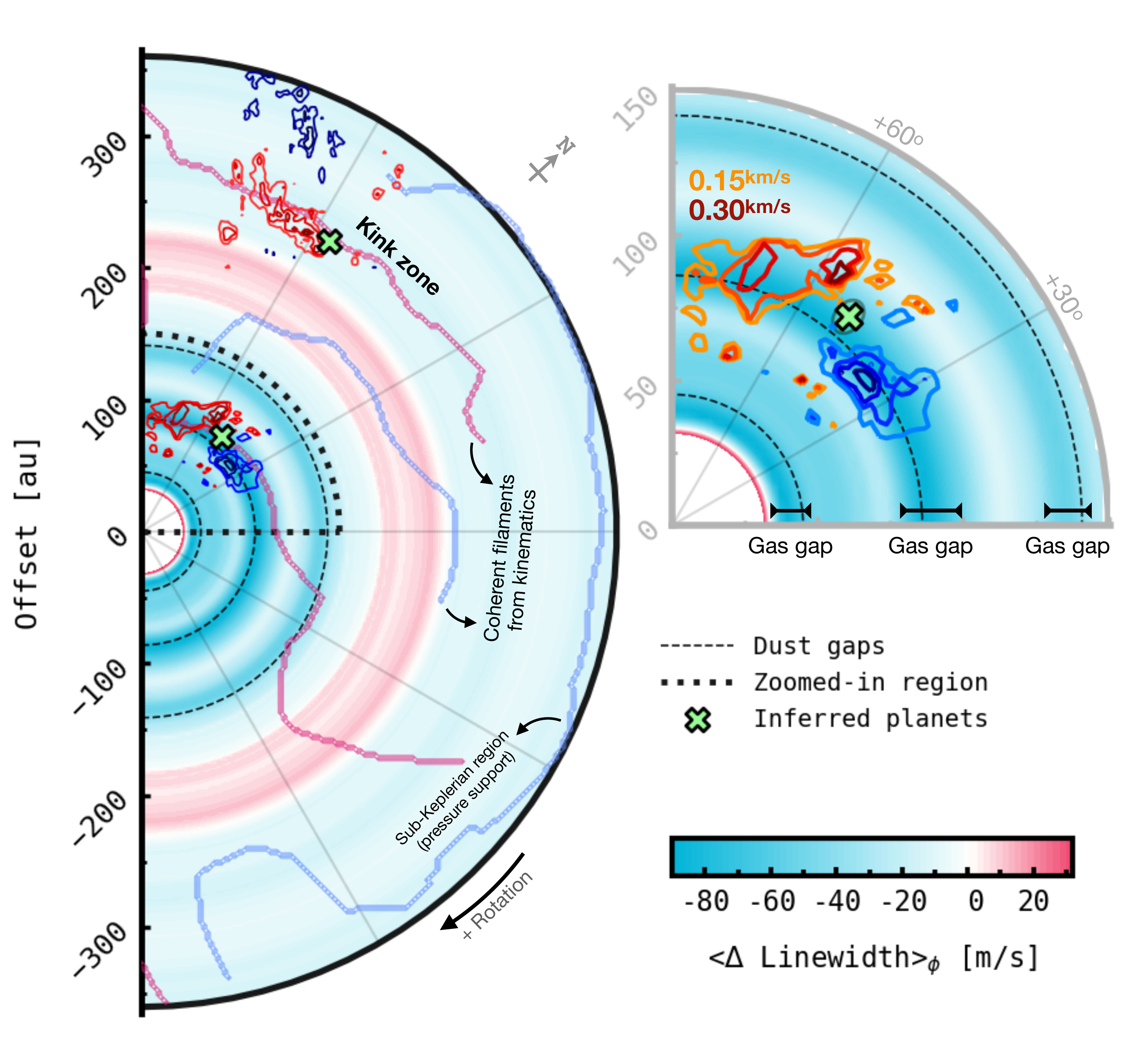} 
     \caption{Summary of the main results of this work. The green crosses mark the inferred location of the P94 and P261 planets, and the blue--red contours, ranging from $\pm$0.15 to $\pm$0.30 in steps of 0.05\,km\,s$^{-1}$, are nearby velocity residuals associated with the detections. The background colours are azimuthally averaged line width residuals to highlight the location of gas gaps. In light red and blue are overlaid the kinematic filamentary structures extracted from centroid velocity residuals on the upper surface of the disc. The quarter circle on the right is a zoom-in around the P94 perturbation, whose nearby velocity contours are this time coloured according to their magnitude. The green ellipse centered on the P94 cross indicates the spatial error of the detection. 
              }
         \label{fig:summary}
\end{figure*} 
%-----------------------------------------------------------------

\subsubsection{Non-detections}

Our algorithm does not detect any localised perturbation around the D141 gap. Assuming that this gap originates from a planet, our rescaled simulations suggest that it should be a low-mass planet of the order of 0.5\,$\Mj$ or less, so that the planet-driven perturbations are blurred with the background $\delta\upsilon\approx0.1$\,km\,s$^{-1}$ velocities (see Fig. \ref{fig:simulations}, right panel). Furthermore, as demonstrated in Fig. 6 of \citetalias{izquierdo+2021}, average line width residuals of only $\Delta L_w\approx-0.05$\,km\,s$^{-1}$ as those around D141 would be compatible with a gas gap opened by a low-mass planet ($<\!1\,\Mj$). On the other hand, as displayed in Fig. \ref{fig:planet_detection}, velocity fluctuations in the D45 gap can be as high as $\delta\upsilon\approx0.45$\,km\,s$^{-1}$, consistent with an embedded giant planet. However, the large azimuthal extent of these perturbations, spanning from $-30^\circ$ to $+60^\circ$ in the disc, prevents the method from detecting any localised signal there, and affects the detection of features beyond. Moreover, D45 is inside the region where the errors in the observed velocities exceed $0.1$\,km\,s$^{-1}$. For these reasons we exclude the D45 gap from the detection analysis. 

\section{Conclusions} \label{sec:conclusions}

We employed the \discminer{} channel-map modelling framework and statistical analysis introduced in \citet{izquierdo+2021} to search for localised velocity perturbations in the disc of \hd{} using \twCOfull{} DSHARP data. Our study aims at retrieving not only radial distance but also azimuth of the localised perturbations, which is a natural step forward in the field. We report the robust detection of two coherent, localised fluctuations possibly driven by two giant planets at $R=94\pm6$\,au, $\phi=50\pm3^\circ$, and $R=261\pm4$\,au, $\phi=57\pm1^\circ$, labeled here as P94 and P261, respectively. The P261 perturbation is in the region of a kink-like feature previously observed by \citet{pinte+2018b} in intensity channel maps, and attributed to an unseen 2\,$\Mj$ planet. The P94 perturbation is consistent with the presence of a 1\,$\Mj$ planet near the centre of the D86 dust gap, which is in turn potentially linked to the radially localised pressure bump reported by \citet{teague+2018a} at $R=83$\,au. The presence of this massive planet could also explain the non-axisymmetric line widths retrieved by our analysis around the D86 gap.

Additionally, we use line profile properties to infer the location of gas gaps and non-axisymmetric substructures in the disc. Based on line width residuals, we detect three gaps centred at $R=38$\,au, $R=88$\,au and $R=136$\,au, compatible with prior radiative transfer models and kinematical measurements of radially localised pressure gradients. On the other hand, the height of the upper emitting surface retrieved by our model at $z/R\approx0.26$ is in good agreement with previous estimates from geometrical and kinematical models. Simultaneously, we provide a model for the lower emitting surface of the disc, which stands at an altitude of $z/R\approx0.2$ above the midplane, and displays brightness temperatures near the CO freeze-out temperature. An illustrative diagram summarising the main findings of this article is presented in Figure \ref{fig:summary}.

\acknowledgments

The authors would like to thank the anonymous referee for their constructive remarks and suggestions which allowed to improve the robustness of the results and the overall quality of the manuscript. This work was partly supported by the Italian Ministero dell Istruzione, Universit\`a e Ricerca through the grant Progetti Premiali 2012 – iALMA (CUP C$52$I$13000140001$), 
by the Deutsche Forschungs-gemeinschaft (DFG, German Research Foundation) - Ref no. FOR $2634$/$1$ TE $1024$/$1$-$1$, 
and by the DFG cluster of excellence Origins (www.origins-cluster.de). 
This project has received funding from the European Union's Horizon 2020 research and innovation programme under the Marie Sklodowska-Curie grant agreement No 823823 (DUSTBUSTERS) and from the European Research Council (ERC) via the ERC Synergy Grant {\em ECOGAL} (grant 855130). SF acknowledges an ESO fellowship. GR acknowledges support from the Netherlands Organisation for Scientific Research (NWO, program number 016.Veni.192.233) and from an STFC Ernest Rutherford Fellowship (grant number ST/T003855/1).

\software{ 
          \textsc{bettermoments} \citep{teague+2018_bettermoments}, 
          \textsc{discminer} \citep{izquierdo+2021}, 
          \textsc{disksurf}
          \citep{teague+2021disksurf}, 
          \textsc{emcee} \citep{foreman+2013}, 
          \textsc{filfinder} \citep{koch+2015}, 
          \textsc{matplotlib} \citep{hunter+2007}, 
          \textsc{scikit-learn} \citep{scikit-learn+2011}.
          }

\setlength{\tabcolsep}{8.0pt} %pad between columns

\begin{table*}
\centering
{\renewcommand{\arraystretch}{1.0}%pad between rows
 \caption{List of attributes considered for the \discminer{} model of \twCOfull{} intensity channel maps from the disc around \hd{}, and the corresponding best-fit parameters. $D_0=100$\,au is a normalisation constant, $z$ is the height above the disc midplane, $R$ is the cylindrical radius,  and $r$ is the spherical radius. The (down-sampled) pixel size of the model is 15.83\,au. PA is the position angle of the semi-major axis of the disc on the redshifted side.}
  \label{table:attributes_parameters}
\begin{tabular}{ llllll } 

\toprule
\toprule
Attribute & Prescription &  \multicolumn{4}{c}{Best-fit parameters for \twCOfull{}} \\
\midrule

Inclination & $i$ & $i=45.71^\circ$ & -- & -- & -- \vspace{0.15cm} \\

Position angle & PA & ${\rm PA}=312.35^\circ$ & -- & -- & -- \vspace{0.15cm} \\

\midrule

Systemic velocity & $\upsilon_{\rm sys}$ & $\upsilon_{\rm sys}=5.77$\,km\,s$^{-1}$ & -- & -- & -- \vspace{0.15cm} \\

Rotation velocity & $\upsilon_k = \sqrt{\frac{GM_\star}{r^3}}R$ & $M_\star =1.97$\,M$_\odot$ & -- & -- & -- \\ \midrule

Upper surface & $z_U = z_0 (R/D_0)^p - z_1 (R/D_0)^q$ & $z_0 = 29.78$\,au & $p=1.21$ & $z_1 = 4.36$\,au & $q=1.98$ \vspace{0.15cm} \\

Lower surface & $z_L = z_0 (R/D_0)^p - z_1 (R/D_0)^q$ & $z_0 = 19.91$\,au & $p=1.09$ & $z_1 = 0.03$\,au & $q=4.18$ \\

\midrule
Peak intensity & $I_p = I_0 (R/D_0)^p (z/D_0)^q$ & $I_0 = 8.23$\,Jy\,pix$^{-1}$ & $p=-4.16$ & $q=3.68$ & --  \vspace{0.15cm} \\

Line width & $L_w = L_{w0} (R/D_0)^p (z/D_0)^q$ & $L_{w0} = 0.08$\,km\,s$^{-1}$ & $p=0.86$ & $q=-1.38$ & -- \vspace{0.15cm} \\ 

Line slope & $L_s = L_{s0} (R/D_0)^p$ & $L_{s0} = 1.85$ & $p=0.21$ & -- & -- \\

\bottomrule

\end{tabular}

  }
\end{table*}

%-----------------------------------------------------------------
\begin{figure*}
   \centering
   \includegraphics[width=0.7\textwidth]{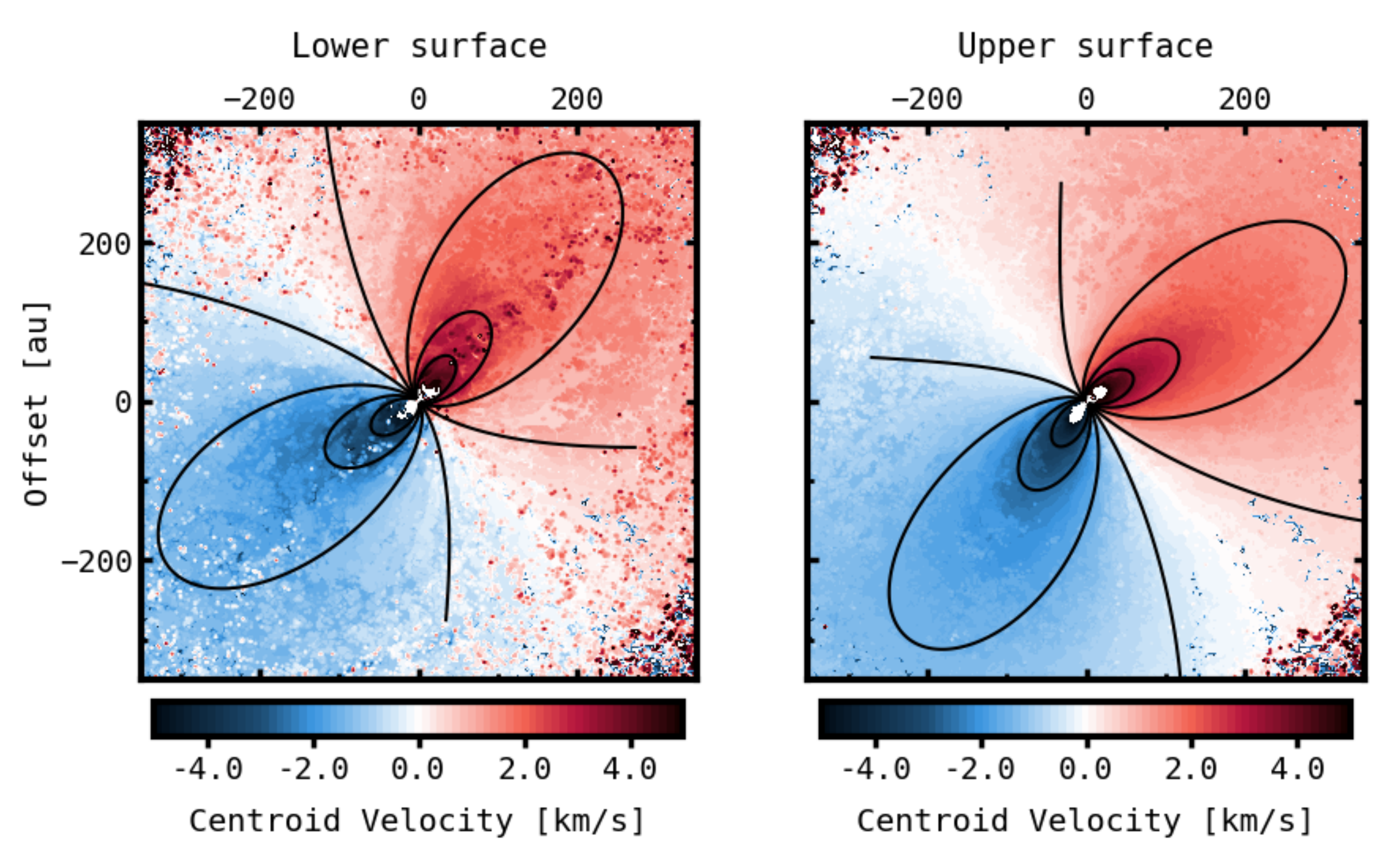} 
      \caption{Empirical reconstruction of centroid velocities from the lower and upper emitting surfaces of the \hd{} disc as observed in \twCOfull{}, shifted to $\upsilon_{\rm sys}=5.77$\,km\,s$^{-1}$. As explained in Sect. \ref{subsec:height}, this model-independent reconstruction consists of fitting double-Bell profiles to the datacube pixels along the velocity axis. For comparison, black contours are line-of-sight velocities from the \discminer{} best-fit model of the lower and upper surfaces of the disc, ranging from $-3.5$ to $3.5$\,km\,s$^{-1}$ in steps of 1.0\,km\,s$^{-1}$.
              }
         \label{fig:centroids_low_up}
\end{figure*} 

%-----------------------------------------------------------------
%-----------------------------------------------------------------
\begin{figure*}
   \centering
   \includegraphics[width=0.7\textwidth]{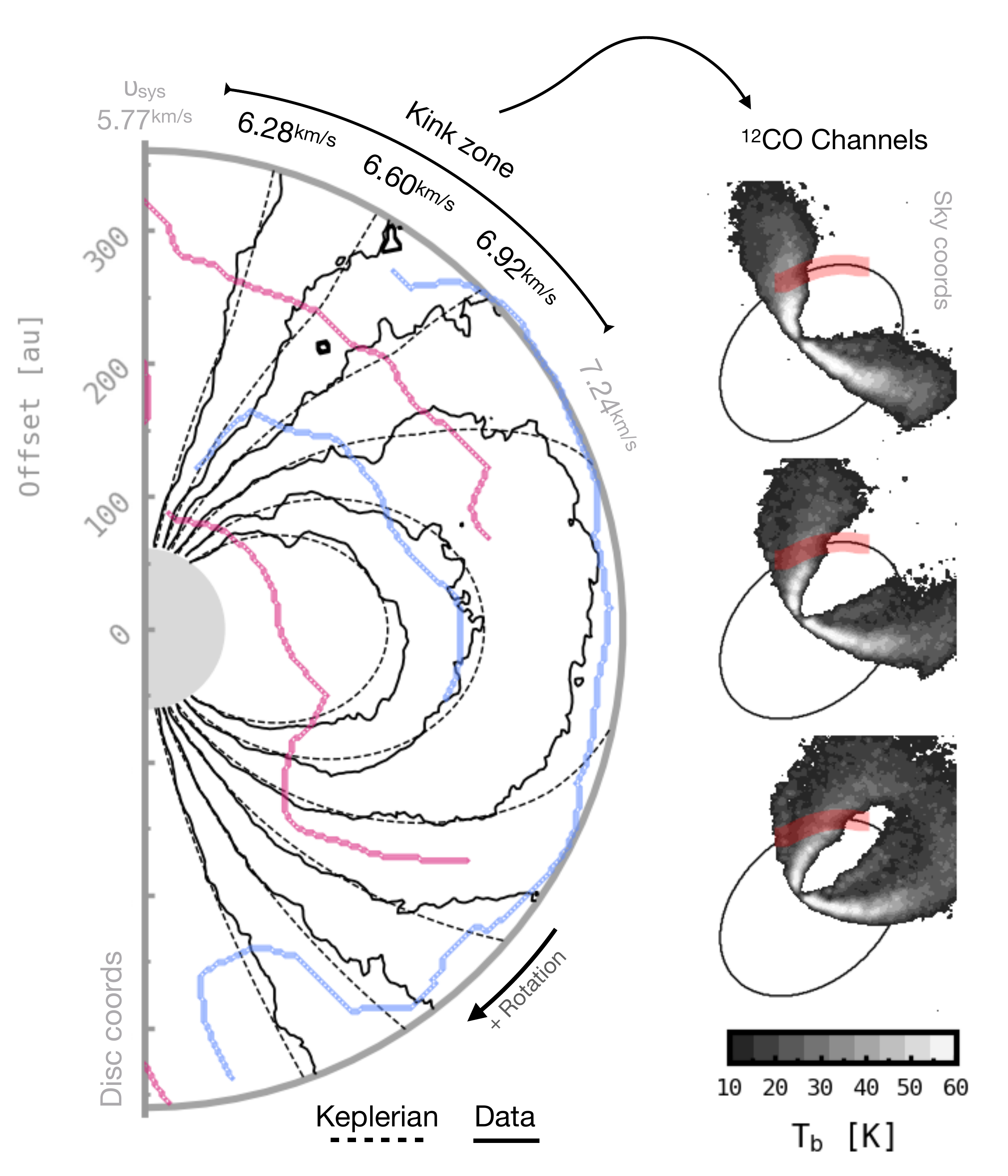} 
      \caption{Illustrating isovelocity contours and deviations from Keplerian rotation related to the presence of coherent structures in the kinematics of \hd{} as seen in \twCOfull{}. The kink-like feature reported by \citet{pinte+2018b} is driven by the long red filament around $R=260$\,au (see also Fig. \ref{fig:residuals_2d}d), spanning from $90^\circ$ to $20^\circ$ azimuths, which corresponds to a range of  $\Delta\upsilon\approx\!1.2$\,km\,s$^{-1}$ in velocity channels. Some of these channels are labelled at the top of the plot within the `Kink zone', and displayed in the right column in sky coordinates. This long kinematic substructure could be closely linked to the localised P261 perturbation and the associated planet candidate reported in Sect. \ref{sec:p261}.
              }
         \label{fig:kink_channels}
\end{figure*} 
%-----------------------------------------------------------------

%-----------------------------------------------------------------
\begin{figure*}
\centering
   \includegraphics[width=0.47\textwidth]{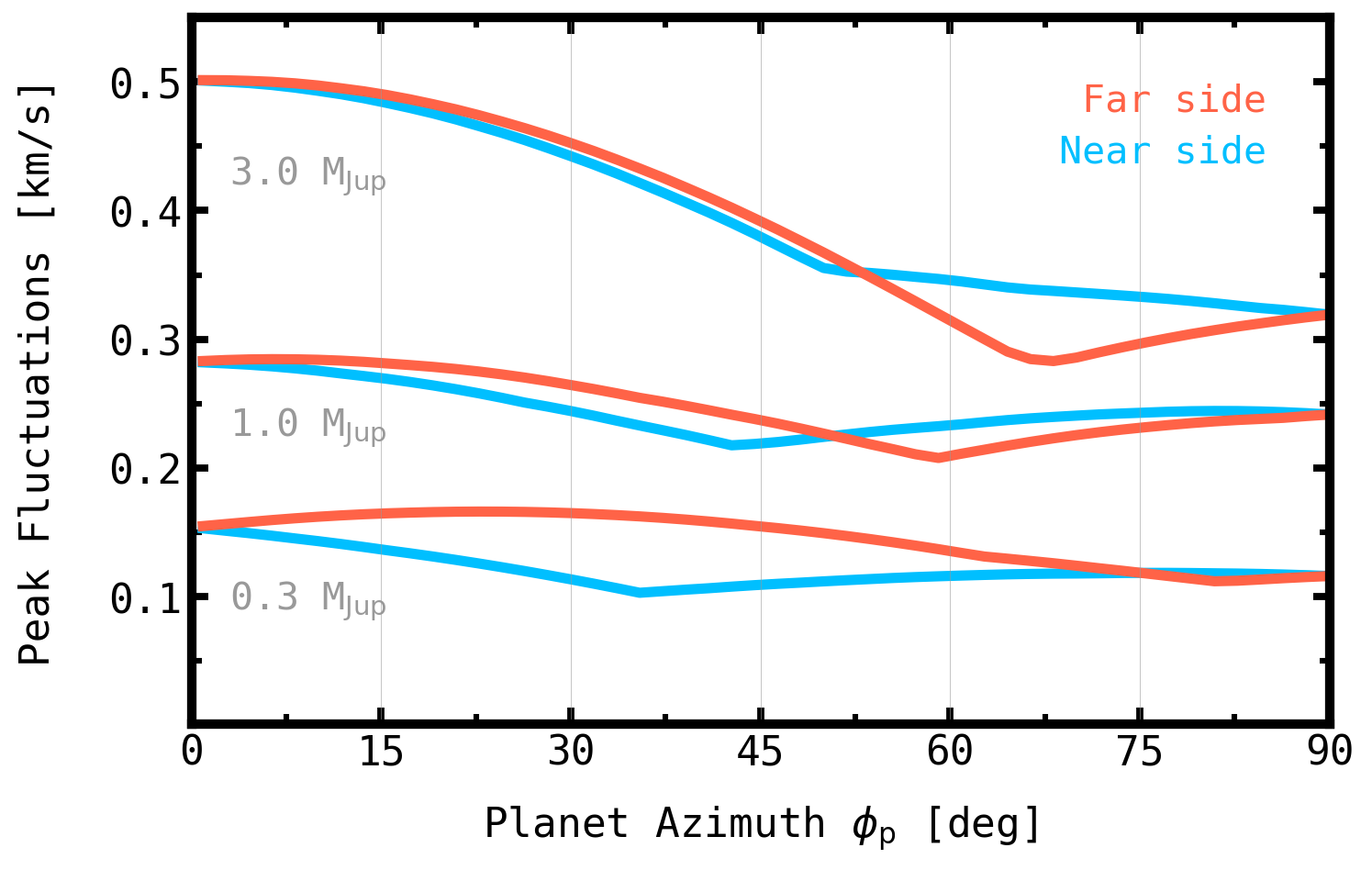} \includegraphics[width=0.47\textwidth]{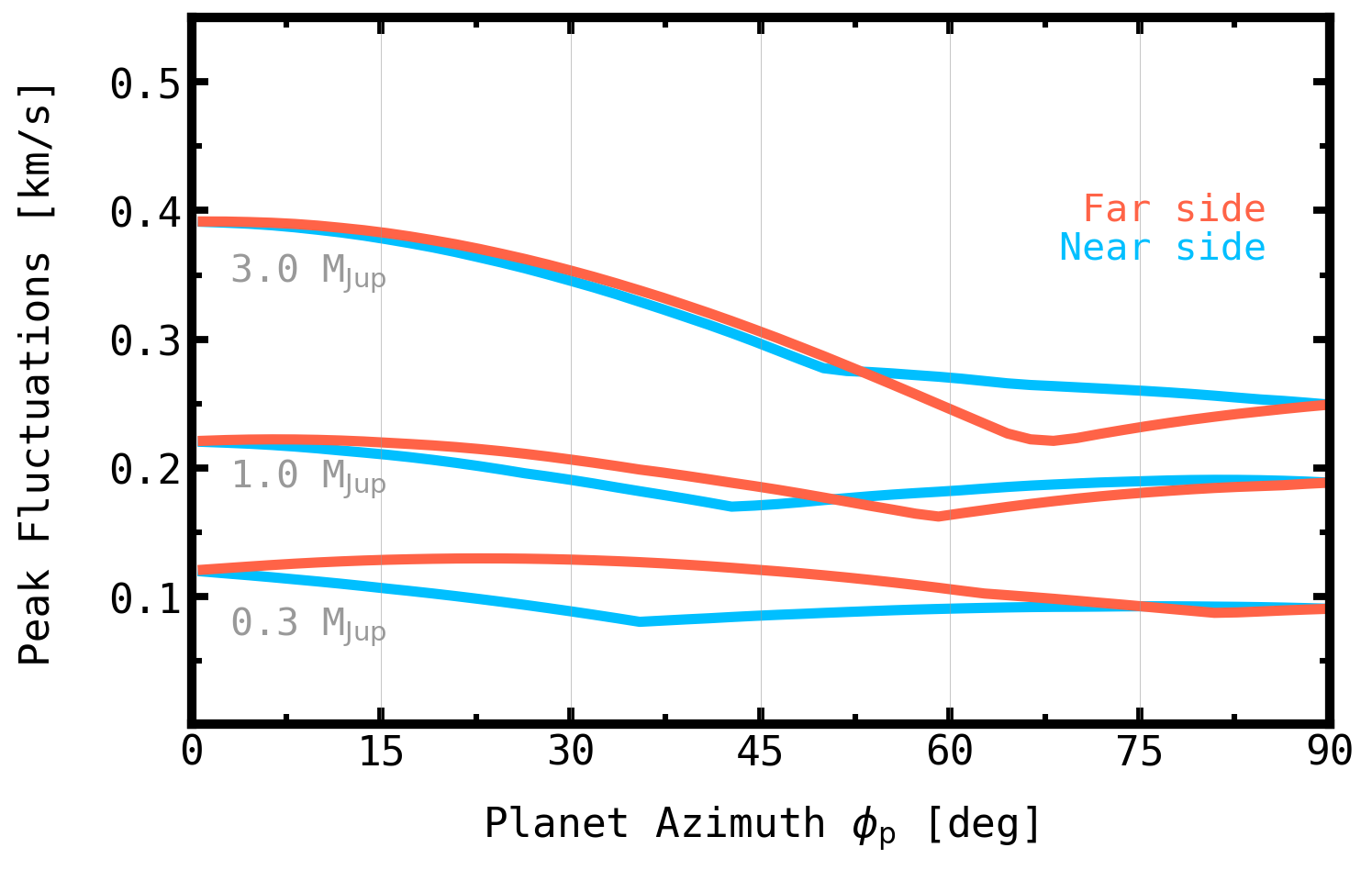} 
      \caption{Simulated intrinsic peak velocity fluctuations as a function of planet azimuth, projected along a line-of-sight parallel to the minor axis of the disc, with an inclination of $45^\circ$. No radiative transfer effects are considered. These simulations are rescaled versions of those presented in \citetalias{izquierdo+2021}, assuming a stellar mass of 2\,M$_\odot$, and three different planet masses at a radius of $R=94$\,au (left) and $R=141$\,au (right). The P94 perturbation detected in this work at $R=94$\,au, $\phi=50^\circ$ on the near side of the disc, is consistent with a $\sim\!1\,\Mj$ planet, whose perturbation is expected to be amplified in folded residual maps due to radiative transfer and projection effects as discussed in Sect. \ref{sec:p94}. The non-detection reported around the D141 gap suggests that an embedded planet at such an orbital radius, if any, should be less massive than half a Jupiter so that its kinematical perturbations go unnoticed with respect to the background fluctuations. 
              }
         \label{fig:simulations}
\end{figure*} 

%-----------------------------------------------------------------
\begin{figure*}
\centering
\includegraphics[width=0.8\textwidth]{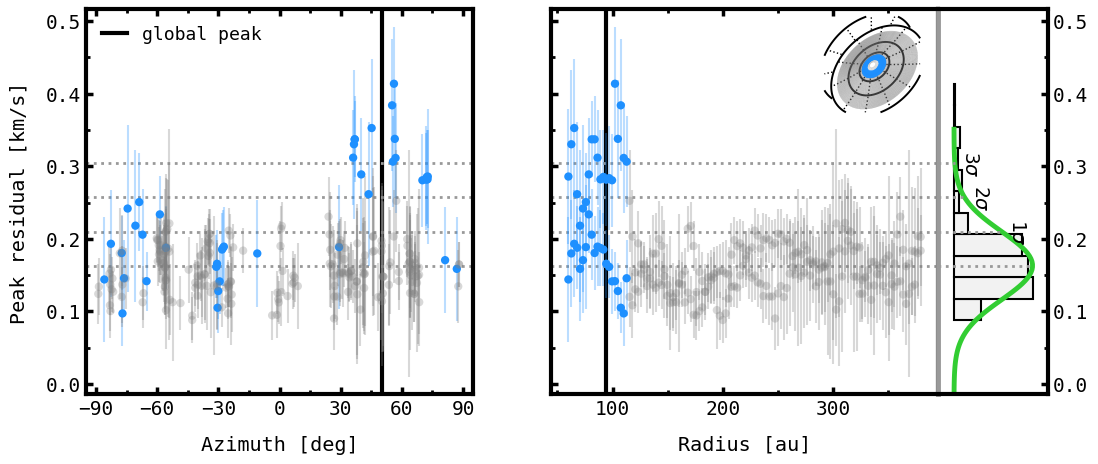}  \includegraphics[width=0.8\textwidth]{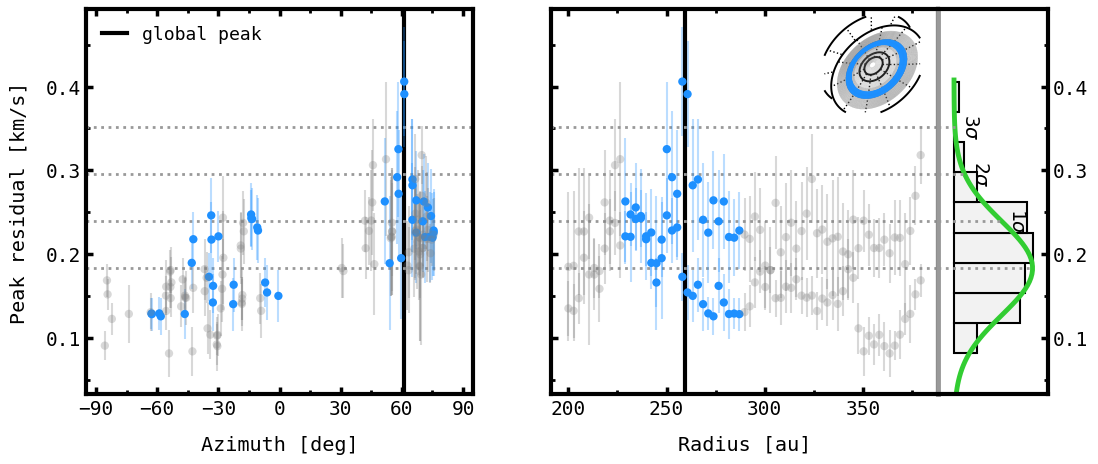} \\
      \caption{ Location of peak velocity residuals in azimuth (left column), and radial distance in the disc (right), obtained from analyses of the upper surface alone (top row) and of the combined contribution of upper and lower surfaces (bottom). Using this kinematical information, the clustering algorithm introduced in Sect. \ref{subsec:planets} detects two localised perturbations, P94 (top) and P261 (bottom), possibly associated with the presence of two giant planets in the disc of \hd{}. Blue circles are residuals extracted from 60\,au wide radial sectors, centred at 86 (top) and 260\,au (bottom). Error bars are computed according to the analytical treatment of uncertainties presented in Appendix \ref{sec:appendix_errors}. The global peaks shown as black lines mark the median location of peak velocity residuals above 3$\sigma$ significance with respect to the background velocities, which nearly follow a normal distribution as displayed in the rightmost subpanels. Note that outliers in the normal distributions are related to the localised perturbations. The mean value of all peak velocity residuals is 0.16 and 0.18\,km\,s$^{-1}$ for the P94 and P261 analyses, respectively, while 1$\sigma$ equals 0.05 and 0.06\,km\,s$^{-1}$, which represents the standard deviation of velocity residuals weighted by their uncertainties.
              }
         \label{fig:folded_azimuthal}
\end{figure*} 

%-----------------------------------------------------------------

\bibliography{references}

\appendix

\section{Analytic propagation of errors} \label{sec:appendix_errors}

To calculate uncertainties in the model attributes and in the derived residual maps, we use analytic formulas for the propagation of errors considering the variance of the posterior distributions of model parameters obtained with the \emcee{} sampler in Sect. \ref{subsec:emcee}, and the variance of the measured line width, velocity and peak intensity maps presented in Sect. \ref{subsec:residuals}.

The validity of this approach is subject to the assumption that the response of the mathematical model $f$, which transforms a set of input variables $\{X_0, X_1,..., X_i,..,X_n\}$ with variances $\sigma_i^2$, into at least one output attribute $Y$ with variance $\sigma_y^2$, is approximately linear within the variable variances. The \discminer{} system is a multi-input multi-output transformation in the sense that it handles multiple input free parameters to model multiple output attributes including peak intensity, line width, emission height, line slope, and rotation velocity, as prescribed in Table \ref{table:attributes_parameters}, as well as intensity channel maps following Eq. \ref{eq:kernel} which depends on the aforementioned attributes.

\begin{figure*}
   \centering
   \includegraphics[width=1.0\textwidth]{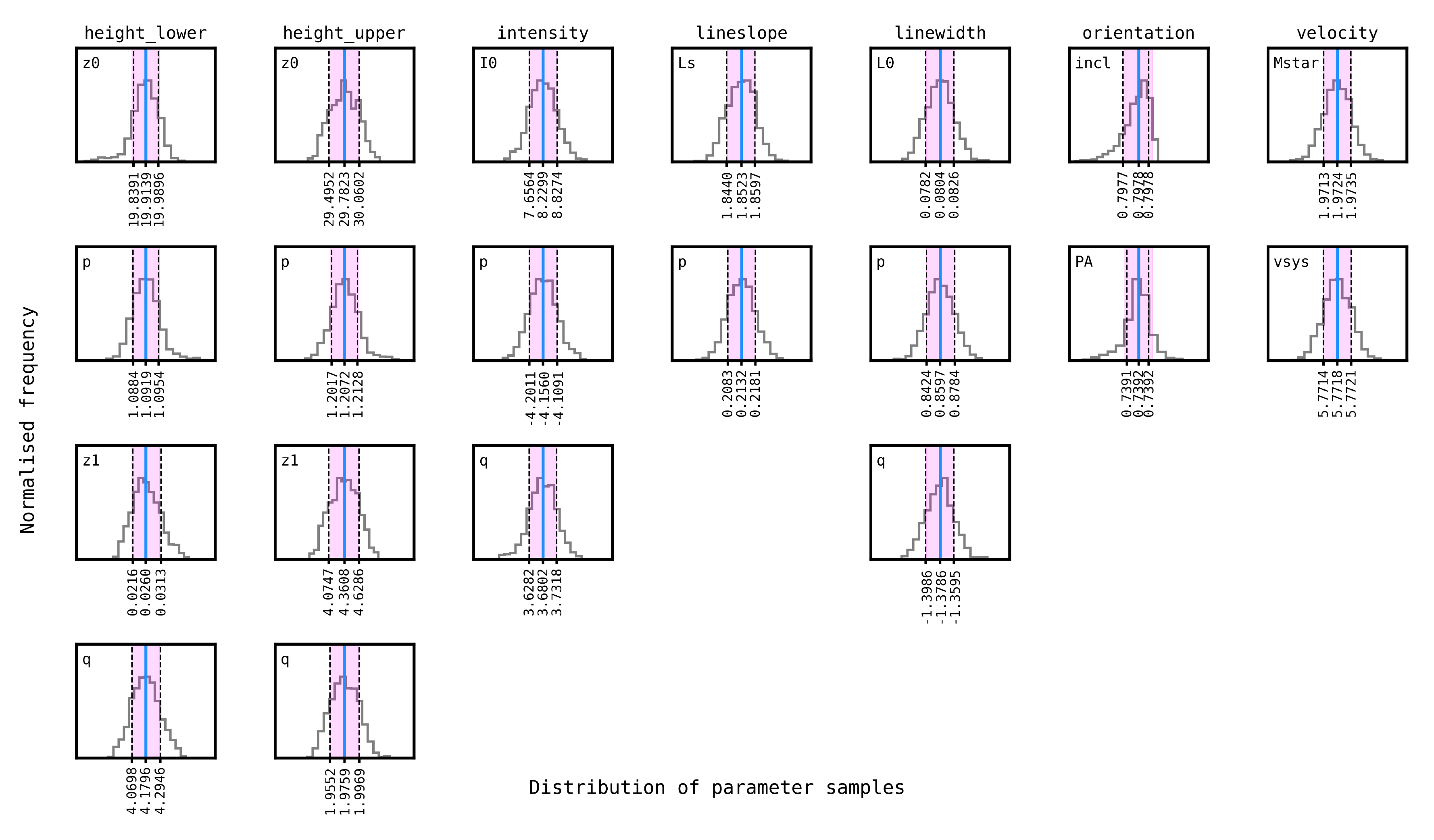}
      \caption{Posterior distributions of \discminer{} model parameters from 256 walkers in the last 5000 steps of a 20000 step run performed by \emcee{}, thinned by half the auto-correlation times of the parameter chains in order to minimise the impact of non-independent samples on the posterior statistics. This run was preceded by a burn-in stage of 3000 steps as specified in Sect. \ref{subsec:emcee}. The blue lines indicate the median of parameters (summarised in Table \ref{table:attributes_parameters}), which are used to generate the best-fit model channel maps. The shades in magenta represent $\pm1$ standard deviations, whereas dashed lines are 15.9 and 84.1 percentiles. The small differences between $1\sigma$ regions and percentiles suggest that the posterior distributions are nearly normal.
              }
         \label{fig:posteriors}
\end{figure*}

A practical approximation of the variance $\sigma_y^2$ of the resulting distribution of the attribute $Y=f(X_0,X_1,...,X_n)$ can be found by writing a first-order Taylor series expansion around the expected value of the input parameters $E[X_0],E[X_1]...,E[X_n]$, and operating from the definition of variance, which gives,

\begin{equation} \label{eq:error_propagation}
    \sigma_y^2 = E[(Y - E[Y])^2] = \sum_{i} \frac{\partial f}{\partial X_i}^2 \sigma_i^2 + \sum_{ij}^{i\neq j}\frac{\partial f}{\partial X_i} \frac{\partial f}{\partial X_j} \sigma_i\sigma_j\rho_{ij},
\end{equation}
where $\rho_{ij} \in [-1,1]$ is the Pearson correlation coefficient between parameter pairs $X_i, X_j$. This analytic formulation has the advantage that the degree of (anti--)correlation between model parameters is accurately considered while keeping processing times short\footnote{We note that even though a direct sampling of the posterior distributions would be more accurate at deriving uncertainties in the residual maps, it does not compensate for the immense computational cost it implies to measure, store and analyse line profile properties from several thousands of model cubes simultaneously.}. The posterior distributions of model parameters and their correlation coefficients are presented in Figures \ref{fig:posteriors} and \ref{fig:error_residuals_2d}, respectively. We note strong (anti--)correlations for several parameters, especially between peak intensity, line width and emission height parameters. Rotation velocity, orientation, and line slope parameters are less dependent on one another.

\begin{figure*}
   \centering
   \includegraphics[width=1.0\textwidth]{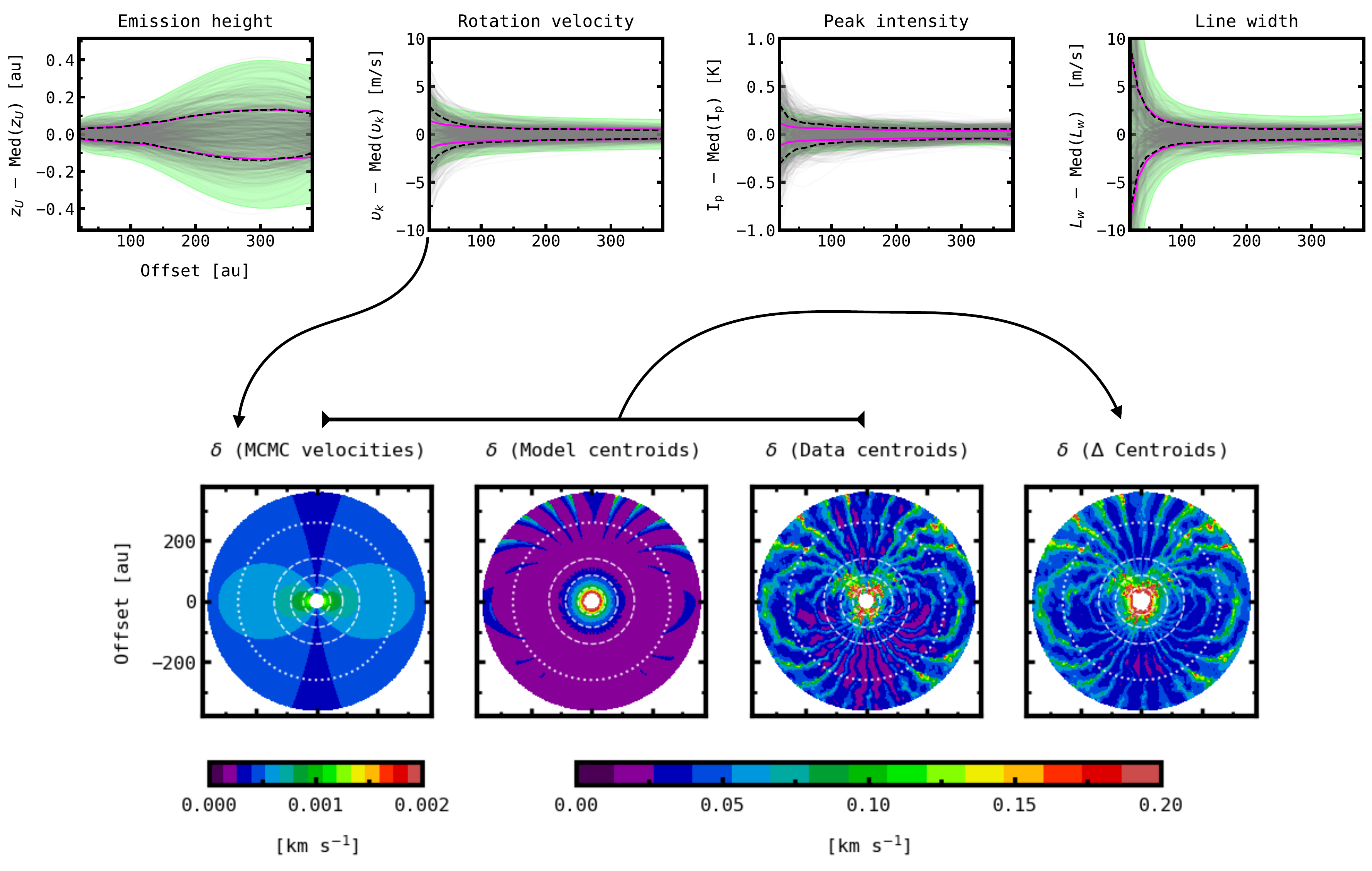}
      \caption{\textit{Top row}: Illustrating how statistical uncertainties of the MCMC search impact the distribution of model attributes. The grey lines are attributes computed for 1000 parameter samples selected randomly from the MCMC posterior distributions. The dashed black lines represent 15.9 and 84.1 percentiles. For comparison, the background shades in green and magenta lines are three and one sigma uncertainties derived analytically using Eq. \ref{eq:error_propagation}, which reproduce the dispersion of model attributes reasonably well. For a better visualization, the model attributes are referred to their median value. Uncertainties in the rotation velocity are projected on the sky plane and computed along $\phi=0^\circ$ azimuth. \textit{Bottom row}: Deprojected uncertainties from the different components involved in the error propagation of centroid velocity residuals. Note that uncertainties in the measurement of data centroids are strongly dominant. The dashed lines indicate the location of the D45, D86 and D141 dust continuum gaps, and the dotted line marks the radial distance of the K260 kink.
              }
         \label{fig:error_attributes}
\end{figure*}

\begin{figure*}
   \centering
   \includegraphics[width=1.0\textwidth]{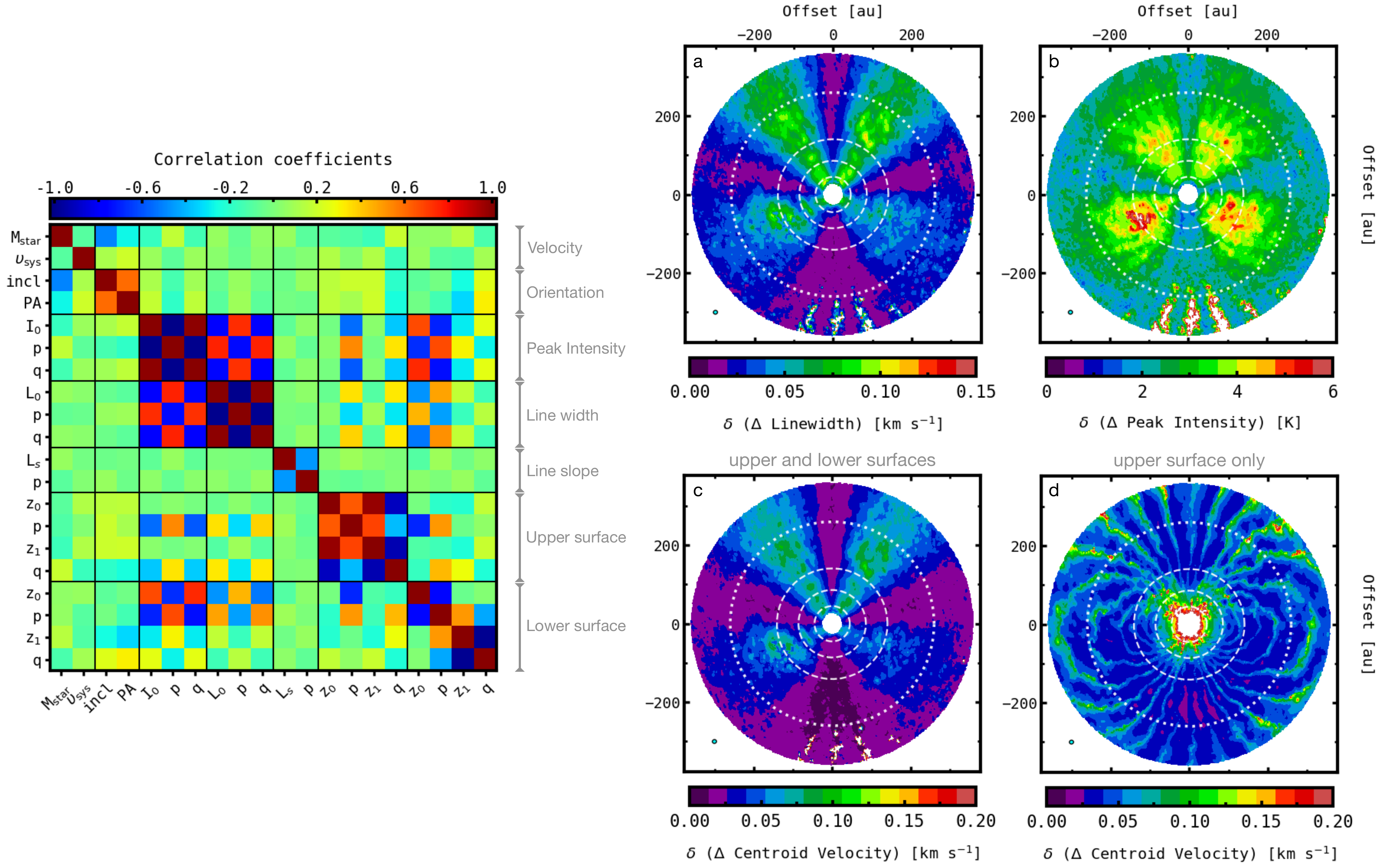}
      \caption{\textit{Left panel}: Pearson correlation coefficients ($\rho_{ij}$) between model parameters. These coefficients are employed to calculate the contribution of crossed derivatives to the output variance of model attributes (see Eq. \ref{eq:error_propagation}). The parameters are grouped according to their corresponding attribute. \textit{Right panels}: Deprojected uncertainties of the residual maps presented in Fig. \ref{fig:residuals_2d}, computed according to the analytical formulation of errors introduced in Appendix \ref{sec:appendix_errors}. The dashed lines indicate the location of the D45, D86 and D141 dust continuum gaps, and the dotted line marks the radial distance of the K260 kink. 
              }
         \label{fig:error_residuals_2d}
\end{figure*}

To verify that the analytical variance of model attributes is a good representation of the statistical uncertainties of the MCMC search, we performed a comparison between three-sigma regions predicted by Eq. \ref{eq:error_propagation} and the model attributes computed with random parameter samples selected from the posterior distributions. In Figure \ref{fig:error_attributes}, we show that both regions are closely equivalent, suggesting that (a) the model response is far from non-linear within at least a few sigma intervals, and (b) the correlations between parameters are reasonably well captured by our analytical treatment.

Once the uncertainties in the model attributes have been computed, they are added to the measurement error of line profile properties to obtain the uncertainty of each model observable, assuming the worst case scenario which is full correlation between both variables. For instance, the uncertainty of the model centroid velocity is simply the sum of the MCMC uncertainty of the rotation velocity on the upper and lower surfaces, and the error in the observed velocity computed through the Gaussian fit, $\sigma_{\upsilon_{c[model]}} = \sigma_{\upsilon_{[lower]}} + \sigma_{\upsilon_{[upper]}} + \sigma_{\upsilon_{obs}}$, or through the quadratic fit from \textsc{bettermoments}, $\sigma_{\upsilon_{c[model]}} = \sigma_{\upsilon_{[upper]}} + \sigma_{\upsilon_{obs}}$.

Finally, the uncertainty in the residual maps is calculated assuming that the measurement errors on the model and the data are independent between each other. Taking the previous example, the uncertainty in centroid velocity residuals is thus given by $\sigma_{\Delta \upsilon _c} = \sqrt{\sigma_{\upsilon_{c[model]}}^2 + \sigma_{\upsilon_{c[data]}}^2}$. It is followed equivalently for line width and peak intensity  residuals. In Figure \ref{fig:error_residuals_2d}, right column, we present the uncertainties derived for the four different residual maps introduced in Fig. \ref{fig:residuals_2d}. We note that the errors in residual maps are dominated by the uncertainties of the data. For instance, the errors of centroid velocities in the data are typically twice as large as the errors of model centroid velocities, of which less than five per cent correspond to statistical uncertainties from the MCMC search. The various components making up the error in centroid velocity residuals are illustrated in the bottom row of Fig. \ref{fig:error_attributes}.

The most visible feature in the error maps of residuals is that those derived from Gaussian fits have a cross pattern following diagonal axes of the disc. This is due to the influence of the lower surface on the line profile which is increasingly prominent as one moves away from the main projected axes of the disc, and so it is the uncertainty of the Gaussian fit properties. It follows reciprocally that the quadratic fit method, which is mainly narrowed to the upper surface of the disc, yields errors that are relatively uniform in azimuth. Nevertheless, errors in centroid velocities measured with Gaussian fits are generally smaller than those from the quadratic fits, for which channelisation effects are already evident. This is why for the analysis of localised velocity perturbations on the upper surface of the disc, presented in Sect. \ref{subsec:planets}, we mask the inner 60\,au where the residual errors are larger than 0.1\,km\,s$^{-1}$ on average.

\section{Supporting figures} \label{sec:supporting_figures}

%-----------------------------------------------------------------
\begin{figure*}
   \centering
   \includegraphics[width=0.8\textwidth]{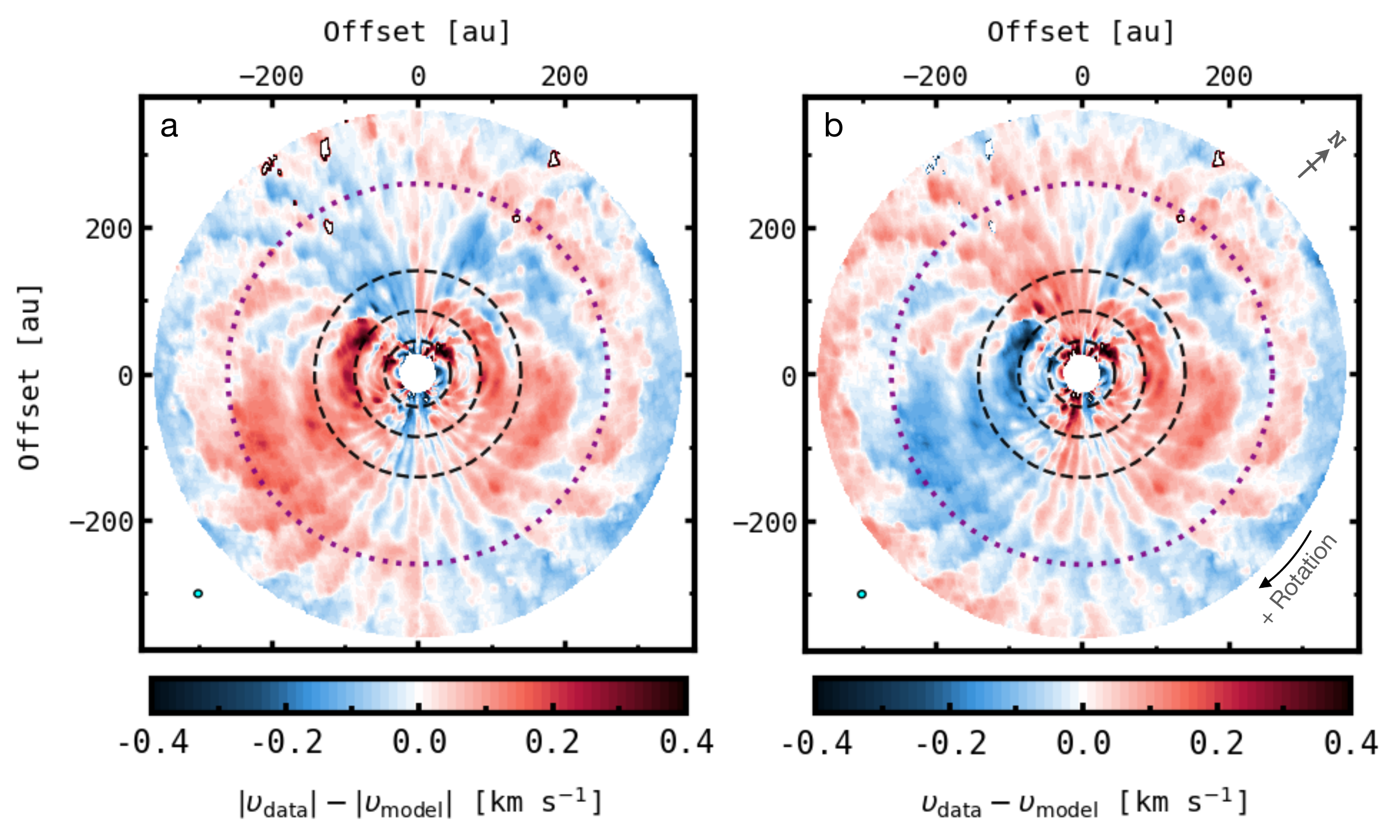} 
      \caption{Comparison between velocity residuals obtained from different subtraction methods. In the left panel, the absolute value of data and model line centroids is taken before subtraction, while in the right panel a direct subtraction is performed. Considering the absolute value can be convenient for visualisation because it makes residuals on the blueshifted side of the disc to switch signs in such a way that sub(super) Keplerian perturbations in the azimuthal component of the velocity, as those expected around gas gaps, appear blue(red). The main signatures of velocity residuals in panel (b) are qualitatively similar to those found by \citet{teague+2021}. Local differences are due to the fact that (1) our model velocities account for the impact of spatial variations in intensity, and (2) we do not consider any kernel to spatially smooth our residual maps. The dashed lines indicate the location of the D45, D86 and D141 dust continuum gaps registered by \citet{isella+2018}. The outer dotted line marks the radial distance of the K260 kink reported by \citet{pinte+2018b}.
              }
         \label{fig:kind_residuals}
\end{figure*} 
%-----------------------------------------------------------------

\end{document}